\documentclass[usenatbib]{mn2e}
\usepackage{natbib}
\usepackage{amsmath}
\usepackage{epsfig}

\def\reff@jnl#1{{\rm#1\/}}

\def\aj{\reff@jnl{AJ}}                  
\def\araa{\reff@jnl{ARA\&A}}            
\def\apj{\reff@jnl{ApJ}}                        
\def\apjl{\reff@jnl{ApJ}}               
\def\apjs{\reff@jnl{ApJS}}              
\def\apss{\reff@jnl{Ap\&SS}}            
\def\aap{\reff@jnl{A\&A}}               
\def\aapr{\reff@jnl{A\&A~Rev.}}         
\def\aaps{\reff@jnl{A\&AS}}             
\def\baas{\reff@jnl{BAAS}}              
\def\jrasc{\reff@jnl{JRASC}}            
\def\memras{\reff@jnl{MmRAS}}           
\def\mnras{\reff@jnl{MNRAS}}            
\def\physrep{\reff@jnl{Phys.Rep.}}
\def\pra{\reff@jnl{Phys.Rev.A}}         
\def\prb{\reff@jnl{Phys.Rev.B}}         
\def\prc{\reff@jnl{Phys.Rev.C}}         
\def\prd{\reff@jnl{Phys.Rev.D}}         
\def\prl{\reff@jnl{Phys.Rev.Lett}}      
\def\pasp{\reff@jnl{PASP}}              
\def\pasj{\reff@jnl{PASJ}}              
\def\skytel{\reff@jnl{S\&T}}            
\def\solphys{\reff@jnl{Solar~Phys.}}    
\def\sovast{\reff@jnl{Soviet~Ast.}}     
\def\ssr{\reff@jnl{Space~Sci.Rev.}}     
\def\nat{\reff@jnl{Nature}}             

\newcommand{\hmpc}{\ensuremath{h^{-1}\mathrm{Mpc}}}
\newcommand{\hkpc}{\ensuremath{h^{-1}\mathrm{kpc}}}
\newcommand{\hMsun}{h^{-1}M_{\odot}}

\newcommand{\ds}{\ensuremath{\Delta\Sigma}}
\newcommand{\Dt}{\ensuremath{\Delta\theta}}
\newcommand{\fdsrot}{\ensuremath{f_{45}\Delta\Sigma}}
\newcommand{\scinv}{\ensuremath{\Sigma_c^{-1}}}

\def\mr{M_r}

\newcommand{\be}{\begin{equation}}
\newcommand{\ee}{\end{equation}}
\newcommand{\bea}{\begin{eqnarray}}
\newcommand{\eea}{\end{eqnarray}}

\title[Halo ellipticity]{Ellipticity of dark matter halos with galaxy-galaxy weak lensing} 

\author[Mandelbaum et. al.]{
Rachel Mandelbaum$^1$\thanks{Electronic address:
    {\tt rmandelb@princeton.edu}},
Christopher M. Hirata$^1$,
Tamara Broderick$^1$,
\newauthor
Uro\v{s} Seljak$^{1,2}$, and
Jonathan Brinkmann$^3$
\\$^1$Department of Physics, Jadwin Hall, Princeton University,
      Princeton NJ 08544, USA
\\$^2$International Centre for Theoretical Physics, Strada Costiera 11,
      34014 Trieste, Italy
\\$^3$Apache Point Observatory, 2001 Apache Point Road,
      Sunspot NM 88349, USA
}
\date{\today}

\begin{document}
\maketitle

\begin{abstract}
We present the results of attempts to detect the ellipticity of dark matter halos using
galaxy-galaxy weak lensing with SDSS data.
We use $2\,020\,256$ galaxies brighter than $r=19$ with photometric redshifts
(divided into colour and luminosity subsamples) as lenses and 
$31\,697\,869$ source galaxies. We search for and identify  
several signal contaminants, which if not removed lead to a spurious 
detection. These include systematic shear that
leads to a slight spurious alignment of lens and source ellipticities,
intrinsic alignments (due to contamination of the source sample
by physically-associated lens source pairs), and anisotropic 
magnification bias. 
We develop methods that
allow us to remove these contaminants to the signal.  
We split the analysis into blue (spiral) and red (elliptical) 
galaxies. Assuming Gaussian errors as in previous work and a power-law
profile, we find
$f_h=e_h/e_g=0.1 \pm 0.06$ for red galaxies and $-0.8 \pm 0.4$ for blue galaxies
using transverse separations of 20--300 \hkpc{}, averaged over luminosity. 
Inclusion of the
more realistic non-Gaussian error distributions and of the NFW density
profile (which predicts much smaller ellipticity of the shear for
scales above the scale radius) yields $0.60 \pm 0.38$ for ellipticals
and $-1.4^{+1.7}_{-2.0}$ for spirals.  While there is no concrete detection of
alignment in either case,
there is a suggestion in the data of a positive alignment in the
brightest lens sample of ellipticals.  
Our results appear to be mildly inconsistent with a
previously reported detection by Hoekstra et al. (2004), but more data
and further tests are needed to clarify whether the discrepancy is
real or a consequence of differences in the lens galaxy samples used
and analysis methods.
\end{abstract}


\section{Introduction}

Dark matter halo ellipticity, a robust prediction of $\Lambda$CDM
according to N-body and hydrodynamic simulations, can in principle be
detected using galaxy-galaxy weak lensing.  In this paper, we
attempt to detect the 
projected ellipticities of the dark matter halos of lens galaxies
using data from the Sloan Digital Sky 
Survey (SDSS).  Besides implications of such a detection for CDM and
theories of hierarchical structure, a positive detection of halo
ellipticity may also be used to constrain
Modified Newtonian Dynamics, or MOND
\citep{1983ApJ...270..365M,1986MNRAS.223..539S,2002ARA&A..40..263S},
an alternative theory of gravity that describes 
rotation curves well without dark matter, but predicts an isotropic
lensing signal around isolated galaxies at scales on which there is no
baryonic matter (though a combination of MOND and dark matter cannot
be ruled out by this test). 

Galaxy-galaxy weak lensing is a useful tool for studying the structure
of dark matter halos on large scales.  Because it is sensitive to all
matter in the lenses, not just baryons, the  signal (averaged over
many lens and source galaxies)  can be detected
well beyond the extent
of the light profiles, without the need for physically associated
tracers such as satellite 
galaxies.  At present, galaxy-galaxy weak lensing has been
well-detected in several surveys (\citealt{1996ApJ...466..623B};
\citealt{1998ApJ...503..531H}; \citealt{2000AJ....120.1198F};
\citealt{2001ApJ...551..643S};  \citealt{2001astro.ph..8013M};
\citealt{2002MNRAS.335..311G}; \citealt{2003MNRAS.340..609H};
\citealt{2004AJ....127.2544S}; \citealt{2004ApJ...606...67H};
\citealt{2005PhRvD..71d3511S}), with
  increasing statistical precision and gradually improving control of
  systematics \citep{2005MNRAS.361.1287M}, and therefore may
also be useful as a tool to study dark matter halo ellipticity.

Ellipticity of dark matter halo profiles has
been predicted in CDM N-body simulations (e.g.,
\citealt{1991ApJ...378..496D}), and observed with
non-lensing methods on scales $<20$ kpc (for a review, see \citealt{1999ASPC..182..393S}).
Simulations typically predict triaxial halos with $b/a > 0.7$ and $c/a
\sim 0.5$, where $b/a$ is the ratio of intermediate to long axis, and
$c/a$ is the ratio of short to long axis.  The resulting mean
projected ellipticity is then 0.3.  It 
should be noted, however, that simulations that include baryons show
reduced halo ellipticity in the inner regions, with
$\Delta(c/a)$ and $\Delta(b/a)\sim 0.1$--0.4
\citep{2004ApJ...611L..73K}; the effect of baryons decreases with radius.  One
measurement of halo ellipticity using galaxy-galaxy 
weak lensing has already been reported.
\cite{2004ApJ...606...67H} argued for a detection of ellipticity by measuring
$f_h=e_h/e_g$, the ratio of the dark matter halo ellipticity to the
ellipticity of the light profile in projection, and found
$f_h=0.77_{-0.21}^{+0.18}$, nominally ruling out an isotropic lensing signal at
the 99.5 per cent confidence level.

A measurement of dark matter halo ellipticity naturally depends on
alignment between the
ellipticity of the light distribution and of the matter distribution.
\cite{2004ApJ...613L..41N} find, using gasdynamical simulations, that
for spiral galaxies on small scales (within the virial radius) the
disk spin axis is aligned with the minor axis of 
the dark matter halo, so when viewing the disks in projection we
should see positive $f_h$.  On larger scales (1-2 \hmpc), when considering the
alignment of the ellipticity of the light of spiral galaxies relative
to large-scale 
structure, the reverse appears to be true, so that $f_h$
should be negative on those scales.  In contrast, comparison of N-body
simulations of satellite distributions with the observed satellite
distribution around the Milky Way \citep{2005MNRAS.363..146L,2005ApJ...629..219Z} suggest
that its disk spin axis is 
aligned with the halo major axis, thus predicting a negative $f_h$ on
all scales.  In short, predictions for spiral galaxies are conflicting.
For elliptical galaxies, the halo and light ellipticities
are expected to be aligned  within the virial radius in the simplest
models, giving positive $f_h$. 

There are, however, a number of difficulties inherent in these measurements.
Some are theoretical; for example, if the dark matter and
light ellipticities are not well aligned, then any tendency for
halo 
ellipticity will be undetectable.  Likewise, as will be shown in \S\ref{SS:elliptheory}, for
some reasonable halo profiles, the ellipticity of the weak
lensing signal is much less than the ellipticity of the projected matter
distribution, so the signature of halo ellipticity in
weak lensing is quite small.  However, several effects are under our
control: the statistics (i.e., finding a large enough sample that halo
ellipticity effects, will be statistically significant); and
PSF-related systematics, which may contaminate
the halo ellipticity measurement, can be studied  and
understood.

The SDSS provides an excellent dataset on which to carry
out these measurements by the above criteria.  With $3\times
10^5$ nearby ($z<~0.25$) 
galaxies with spectroscopy that can serve as lenses, or eight
times as many bright ($r<19$) lenses if we do not require
spectroscopy, and roughly $3\times 10^7$ fainter 
galaxies to serve as sources, the SDSS should provide the statistical
power to measure these effects.   
Also, as will be
described in \S\ref{SS:systematics}, we have the tools to isolate the
effects of PSF systematics on the halo ellipticity measurement.

We begin by outlining the weak lensing formalism and the models
used for the ellipticities of the dark matter halo 
in \S\ref{S:theory}.  Next, in \S\ref{S:data}, we
describe the data used for these measurements and systematic effects
that may bias the measurements.  \S\ref{S:analysis} outlines the
method of analysis that we use to relate results from the data to our
models in \S\ref{S:theory}.  Results are presented in
\S\ref{S:results}, and the implications and suggestions for future
work are in \S\ref{S:discussion}.

Here we note the cosmological model and units adopted for this work.
Pair separations are measured in transverse comoving \hkpc{} using the
angular diameter distance (in a flat $\Lambda$CDM cosmology with $\Omega_m=0.3$,
$\Omega_{\Lambda}=0.7$) at the lens redshift.  In the units used, our
results are independent of $H_0$.  

\section{Theory}\label{S:theory}

\subsection{Galaxy-galaxy weak lensing formalism}\label{SS:ggwl}

In this paper, we use the notation and approach to the lensing signal
computation from \cite{2005MNRAS.361.1287M}, 
hereinafter M05.  In this section we briefly mention the main points
of the analysis.

Galaxy-galaxy weak gravitational lensing is sensitive to the
differential surface density $\ds(r)=\overline{\Sigma}(<r)-\Sigma(r)$
(for an axisymmetric halo) averaged over many lenses.  This surface density can be related to a
product of a tangential shear term and a redshift-dependent term,
$\ds=\gamma_t\Sigma_c$, where the critical surface density is defined
in comoving coordinates via
\be
\Sigma_c = \frac{c^2}{4\pi G} \frac{D_s}{D_{ls}D_l(1+z_l)^2}
\ee  
where $D$ is the angular diameter distance, and $l$ and $s$ refer to
the lens and source, respectively. 

The tangential shear is computed by finding the average value of
tangential ellipticity, but the two are related by a factor of 2
according to our definitions; furthermore, we must divide by a shear
responsivity factor.  We write
\be
\ds = \frac{\sum_i w_i (e_t \Sigma_c)_i}{2{\cal R}\sum_i w_i}
\ee
where the summation is over lens-source pairs, ${\cal R}$ is the shear
responsivity which describes the effects of a shear on the tangential
ellipticities \citep{2002AJ....123..583B}, and the weights are defined
by 
\be
w_i = \frac{\Sigma_{c,i}^{-2}}{\sigma_{e_t,i}^2+\sigma_{SN}^2},
\ee
where the weighting is by inverse measurement noise plus shape noise.
Since typical rms ellipticities are of order 0.36 and typical shears
are of order $10^{-3}$, we must average over millions of lens-source
pairs to get reasonable signal to noise.  The shear responsivity and
weighting scheme are discussed in more detail in M05.  Once these sums
have been accumulated, we must subtract off the signal measured around
random points, and boost by the number of pairs relative to the number
around random points to account for dilution of the signal by
physically-associated pairs.

\subsection{Our Model}
Here we describe the model used in this paper for halo
ellipticity. 
This formalism was developed by \cite{2000ApJ...538L.113N} and applied
to the elliptical isothermal sphere density profile.

We attempt to determine a parameter $f$ which is related to the
ellipticity of $\Delta\Sigma$, the differential dark matter surface
density of galaxies.  We model $\Delta\Sigma$ as having an isotropic
component $\Delta\Sigma_{iso}$ and some azimuthal variation as
follows:
\be\label{E:model}
\ds_{model}(r) = \ds_{iso}(r)\left[ 1 + 2\,f\,e_g\,\cos{(2\Dt)}\right]
\ee
where $e_g$ is the observed ellipticity of the light distribution of
the lens and $\Dt$ is the angle from the lens major axis.
We can determine $\ds_{iso}$ and $f$ by minimizing
\be
\chi^2 = \frac{\sum_i w_i (\ds_{model}-\ds_i)}{\sum_i w_i}
\ee
where the summation is over lens-source pairs denoted by $i$ with
weights described in \S\ref{SS:ggwl} and $\ds_i = e_t \Sigma_c$ for the
pair $i$.  By minimizing with respect to $f$ and $\ds_{iso}$,
we obtain the following joint solution for the two 
quantities\footnote{This results in minimization if $\sum_i w_i 
e_{g,i}\cos(2\Delta\theta_i)$ is negligibly small, as would be expected 
if the sources are isotropically distributed relative to the lens major 
axis.  The validity of this approximation is extensively discussed in 
\S\ref{sss:aniso}.}:
\begin{align}\label{E:sums}
\ds_{iso}(r) &= \frac{\sum_i w_i \ds_i}{\sum_i w_i} \\
f\ds_{iso}(r) &= \frac{\sum_i w_i \ds_i e_{g,i}
  \cos{(2\Dt_i)}}{2\sum_i w_i e_{g,i}^2\cos^2{(2\Dt_i)}}\notag
\end{align}
The solution for $\ds_{iso}(r)$ is thus as expected, the usual
averaging of tangential ellipticity over all pairs, and the
calculation of $f\ds_{iso}(r)$ is a summation weighted by
$e_g\cos{(2\Dt)}$, which will pick out any tendency for
$\ds$ to be larger (smaller) along the major axis of the projected light
distribution, yielding a positive (negative) value for
$f\ds_{iso}(r)$.  This procedure can be used even if $f$
is not a constant with radius, allowing us to measure $f(r)\ds_{iso}(r)$.

We also will find it necessary to calculate $\ds_{45}$, the 
signal with source ellipticities rotated by $\pi/4$.  While this
quantity must average out to zero by symmetry when averaged
over $\Dt$, we will see in \S\ref{SS:elliptheory} that it can have a
$\sin{(2\Dt)}$ dependence.  If we also rotate the lens ellipticities in
the same direction as the source ellipticities, that $\sin{(2\Dt)}$
dependence becomes a
$\cos{2\Dt}$ dependence as for the tangential shear, i.e.
\be
\Delta\Sigma_{45,model}(r) = \Delta\Sigma_{iso}(r)\left[2f_{45}e_g\cos{(2(\Dt+\pi/4))}\right]
\ee
so we can write
\be\label{E:sums45}
f_{45}\ds_{iso}(r) = \frac{\sum_i w_i \ds_{i,45} e_{g,i}
  \cos{[2(\Dt_i+\pi/4)]}}{2\sum_i w_i e_{g,i}^2\cos^2{[2(\Dt_i+\pi/4)]}}.
\ee

Once these sums have been calculated, the usual prescription must be
followed as already described: divide by $2{\cal R}$, subtract off
the random catalog signal from $\ds_{iso}$, and multiply by the boost factor $B(r)$, the ratio
of lens-source pairs using real lenses to the number using random
points as lenses, which goes to one in the limit of no physically
associated sources included in the sample, and is larger than that at
small transverse separations. Multiplying by this factor
corrects the signal for dilution due to these non-lensed galaxies
included in the source sample.

There are a number of difficulties to consider when attempting to
detect dark matter halo ellipticity that complicate this simple model
we have presented here,
in particular difficulties in getting adequate signal to noise, the question of
whether the light and mass ellipticities are actually aligned,
choosing radial ranges to best carry out this measurement, and
systematics that obscure the measurement.  All of these issues will be
addressed in the sections that follow.

\subsection{Connection to theory and other work}\label{SS:elliptheory}

In order to facilitate a comparison of our results with theory and
with observational results from
\cite{2004ApJ...606...67H}, we must relate the quantities we measure
to theoretical predictions for various halo density profiles.

We consider several different types of profiles.  The simplest is the general
power-law profile, for which we derive analytic expressions for $f$
correct to first order in the ellipticity.  More realistic profiles,
particularly the Truncated Isothermal Sphere (TIS)
 and the Navarro-Frenk-White 
(NFW) profile, are also considered, with
numerical results given.  In all cases, we introduce ellipticity by
replacing $r$ in the isotropic form for
$\kappa(r)$ with an elliptical coordinate
to compute (analytically or numerically)  the potential and,
ultimately, the shears
$\gamma_t(r)$ and $\gamma_{45}(r)$. We find that the degree to which the
ellipticity of $\kappa$ is expressed as ellipticity of $\ds$ depends
on the shape of the profile.

For a dark matter halo with power-law space density profile $\rho(r) \propto
r^{-(1+\alpha)}$ and ellipticity
independent of radius $e_h$, oriented with major axis along $\theta=0$
and $\pi$ for simplicity, we can write the surface density as
\begin{align}\label{E:kappamodel}
\kappa(r,\theta) &= A\left[r\left(
  1-\frac{e_h}{2}\cos{(2\theta)}\right)\right]^{-\alpha} \\
&\cong Ar^{-\alpha}\left[ 1+\frac{e_h\alpha}{2}\cos{(2\theta)}\right].
\end{align}
This definition ensures that for major axis $a$ and minor axis $b$,
the relationship $e_h=(a^2-b^2)/(a^2+b^2)$ holds.  The next order
term, proportional to $\cos^2{(2\theta)}$, has a coefficient that is
smaller than that of the $\cos{(2\theta)}$ term by a factor of
$e_h(1+\alpha)/4$, which for typical values $\alpha\sim 1$ and
$e_h\sim 0.3$ is roughly 0.15.  Consequently we will ignore all terms
of order higher than $\cos{2\theta}$ for the remainder of this
calculation.

For an axisymmetric (around the line of sight) matter distribution, we 
could simply obtain \ds{} by
finding the average value of $\kappa$ within radius $r$ and using
$\ds=\Sigma_c\left[\overline{\kappa}(<r)-\kappa(r)\right]$.  However,
for non-axisymmetric halos, we must do the full calculation, first
determining the lensing potential corresponding to the surface density via
\be
\kappa = \frac{1}{2}\nabla^2 \Phi =
\frac{1}{2}\left[r^{-1}\frac{\partial}{\partial
    r}\left(r\frac{\partial\Phi}{\partial r}\right) +
  r^{-2}\frac{\partial^2\Phi}{\partial\theta^2}\right]. 
\ee 
With an ansatz of $\Phi=\sum_m \Phi_0(m)
r^{-\beta(m)}\cos{(m\theta)}$, our first order approximation
(Eq.~\ref{E:kappamodel}) gives a potential of
\be
\Phi(r,\theta) = \frac{2A}{(\alpha-2)^2}r^{2-\alpha} \left[ 1 + \frac{B(\alpha-2)^2}{\alpha(\alpha-4)}\cos{(2\theta)}\right]
\ee
where $B=e_h \alpha/2$.  Since this expression diverges for
$\alpha=2$ ($\rho\propto r^{-3}$), it cannot be used to derive
expressions for the shear and $f$ for NFW profiles well
beyond the scale radius.

We can then use the potential to determine $\gamma_t$ and
$\gamma_{45}$ via 
\begin{align}\label{E:theorgamma}
\gamma_t(r,\theta) &= \frac{1}{2}\left(-\frac{\partial^2\Phi}{\partial
  r^2}+r^{-1}\frac{\partial\Phi}{\partial
  r}+r^{-2}\frac{d^2\Phi}{d\theta^2}\right) = \kappa -
\frac{d^2\Phi}{dr^2} \\
\gamma_{45}(r,\theta) &=
-r^{-1}\frac{\partial^2\Phi}{\partial\theta\partial r} + r^{-2}\frac{\partial\Phi}{\partial\theta}\notag
\end{align}

For a general power-law potential, we therefore find
\begin{align}\label{E:gammapowerlaw}
\gamma_t &=
\frac{\alpha}{2-\alpha}Ar^{-\alpha}\left[1+\frac{(\alpha-2)(\alpha^2-2\alpha+4)B}{\alpha^2(\alpha-4)}\cos{(2\theta)}
  \right] \\
\gamma_{45} &= \frac{\alpha}{2-\alpha}Ar^{-\alpha} \left[ \frac{4(2-\alpha)(1-\alpha)B}{\alpha^2(\alpha-4)}\right]\sin{(2\theta)}\notag
\end{align}
We can then make the identification $\ds_{iso}(r) = \Sigma_c\alpha
Ar^{-\alpha}/(2-\alpha)$ to relate back to our model in Eq.~\ref{E:model}.

In the special case of an elliptical SIS ($\alpha=1$), we have
$d^2\Phi/dr^2=0$, $\gamma_t=\kappa$, and therefore
\begin{align}\label{E:SISshear}
\gamma_t &= Ar^{-1}\left[ 1 + \frac{e_h}{2}\cos{(2\theta)}\right] \\
\gamma_{45} &= 0\notag
\end{align}
Comparison against our model in Eq.~\ref{E:model} shows that our
$f=e_h/4e_g$, so that for $e_h=e_g$ ($f_h=1$) we expect $f=0.25$.  Dividing the
measured $f$ by the predicted SIS value $f/f_h=0.25$ then allows us to
compute the measured $f_h$ from our $f$ value assuming the SIS density profile.

More generally, for a power-law non-SIS model, we find
\be
f = \frac{f_h}{4}\left[ \frac{(\alpha-2)(\alpha^2-2\alpha+4)}{\alpha(\alpha-4)} \right]
\ee
so for $\alpha=0.8$--0.9 (a typical value in the actual data at small
transverse separations) $f\sim 0.33f_h$.

We are also interested in computing $f_{45}$ for these profiles.
By comparison with previous equations, we find
\be
f_{45} =
f_h\left[\frac{(2-\alpha)(1-\alpha)}{\alpha(\alpha-4)}\right]
= \frac{4(\alpha-1)}{\alpha^2-2\alpha+4}f
\ee
Consequently the expected signal with lenses and sources rotated by
$\pi/4$ can be predicted relative to the unrotated signal, and the
comparison of the two is an important consistency check.  For
$\alpha\sim 0.8$--0.9 as is found in the data, $f_{45}/f\sim -0.26$
to $-0.13$, i.e. the rotated signal is smaller than the
unrotated one and with opposite sign.  As will be described in
\S\ref{SS:systematics}, to eliminate contributions from systematics to
$f\ds$, we actually will measure $f-f_{45}$, which for the general
power-law profile is
\be\label{E:fminusf45}
f-f_{45} = \frac{(\alpha-2)(\alpha-4)}{\alpha^2-2\alpha+4}f = \frac{f_h}{4}\left[\frac{(\alpha-2)^2}{\alpha}\right]
\ee

We also consider non-power law density profiles. The first
such profile that we will consider is the TIS
\citep{1996ApJ...466..623B}, which takes the form
\be
\rho(r) \propto \frac{1}{r^2(r_s^2+r^2)}
\ee
(note that what we call $r_s$ here is often called $s$; we use $r_s$
for simplicity of notation since it also appears in the NFW profile).
This profile yields $\kappa \propto r^{-1}$ for $r\ll r_s$ and $\propto
r^{-3}$ for $r \gg r_s$.  While analytic expressions can be derived
for the shear in the spherical case, numerical integration must be
used when $r$ is replaced by an elliptical coordinate.  Fortunately,
$\kappa$ and $\gamma$ can be computed for this model using {\sc
  gravlens}\footnote{The latest
  version may be found via the link at {\slshape
    http://cfa-www.harvard.edu/castles/}.} \citep{2001astro.ph..2340K}
via subtraction of the shear from an elliptical $\kappa\propto (r^2+s^2)^{-1/2}$ model from a
$\kappa\propto r^{-1}$ model (due to linearity of the shear), with $r$ being
replaced by an elliptical coordinate.  We can use the predictions for
those quantities to 
compute $f-f_{45}$ by finding, for a profile with ellipticity $e$,
\begin{align}
\Delta(r) &\equiv \gamma_t(r,\theta=0) - \gamma_t(r,\theta=\pi/2) \notag\\ 
\Delta_{45}(r) &\equiv \gamma_{45}(r,\theta=\pi/4) - \gamma_{45}(r,\theta=3\pi/4)
\notag\\ 
T(r) &\equiv \gamma_t(r,\theta=0) + \gamma_t(r,\theta=\pi/2) \\
f(r) &= \frac{\Delta(r)}{2eT(r)} \notag \\
f_{45}(r) &= \frac{\Delta_{45}(r)}{2eT(r)} \notag
\end{align}

\begin{figure}
\includegraphics[width=3in,angle=0]{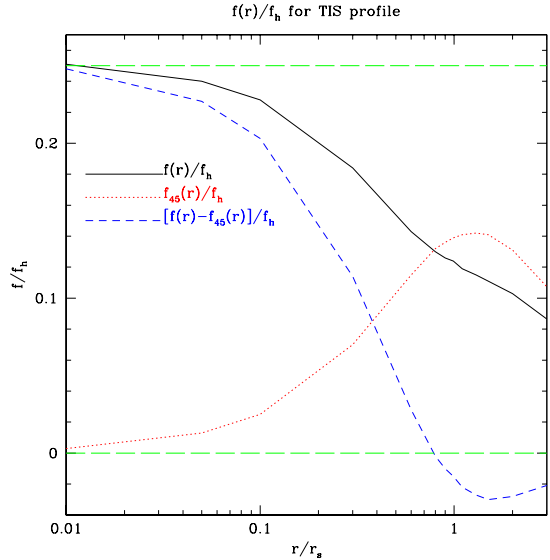}
\caption{\label{F:fradTIS} Plot of predicted $f(r)/f_h$ and $f_{45}(r)/f_h$
  for an elliptical TIS density
  profile dark matter halo; the maximum radius shown, $3r_s$, is
  larger than the scales used in this paper.  Horizontal lines
  indicate the SIS predictions $f/f_h=0.25$ and $f_{45}/f_h=0$.}  
\end{figure}

The plot in Fig.~\ref{F:fradTIS} shows the results for $f$, $f_{45}$,
and $f-f_{45}$  as a
function of $r/r_s$ for the TIS over a wide range of scales. (This plot was
computed using $e=0.02$, for which the variation of the shear with
azimuthal angle is dominated by the first-order $\cos{(2\theta)}$ term;
at more realistic values like $e=0.3$, higher order terms may
contribute as much as 18 per cent of the value of this first order
term.  However, as will be shown, since our results include
statistical uncertainty larger than this value, we will henceforth
neglect higher order terms.)   Notice that $f$ for the TIS is
 a declining function of transverse separation, and for the smallest
 scales shown on the plot approaches the SIS prediction $f/f_h=0.25$.
Because $f_{45}$ is so large, $f-f_{45}$ is quite close to zero for
$r\sim 0.6r_s$ to the largest scales shown on the plot, $r\sim
3r_s$.  

\begin{figure}
\includegraphics[width=3in,angle=0]{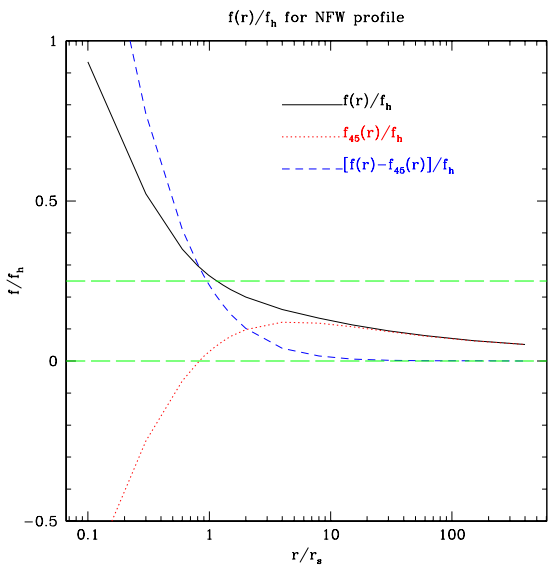}
\caption{\label{F:fradNFW} Plot of predicted $f(r)/f_h$ and $f_{45}(r)/f_h$
  for an elliptical NFW density
  profile dark matter halo.  Horizontal lines
  indicate the SIS predictions $f/f_h=0.25$ and $f_{45}/f_h=0$.}
\end{figure}

We also consider 
the NFW profile \citep{1996ApJ...462..563N}, which takes the form
\be
\rho = \frac{\rho_s}{(r/r_s)(1+r/r_s)^2},
\ee
where $r_s=r_v/c$ (concentration parameter). While the shear for the 
spherical NFW model can be computed
analytically~\citep{2000ApJ...534...34W}, 
the additional complication of ellipticity necessitates the use of
numerical integration to get the shear.  So, we again use $\gamma$
from {\sc gravlens} to compute $f$ and $f_{45}$.  A plot 
of $f(r)$ for this model is shown in Fig.~\ref{F:fradNFW}.  As shown,
the NFW model gives decreasing $f(r)$ as for the TIS model.  While at
$r=r_s$, the NFW model gives $f/f_h\sim 0.3$ (slightly larger than the
SIS) and $f_{45}\ll f$, $f$ decreases at larger radii and $f_{45}$
increases so that $f-f_{45}$ is quite small for $r$
larger than about $2r_s$.

For the sake of clarity,  Table~\ref{T:defs} summarizes the definitions
of all ellipticity-related parameters used in this paper.
\begin{table}
\caption{\label{T:defs} Definitions of all measures of projected ellipticities
  (of dark matter halo and source distributions) used in this paper.}
\begin{tabular}{cl}
\hline\hline
$f$ & Ellipticity of $\ds= \ds_{iso}[1+2fe_g\cos{(2\Dt)}]$ \\
$f_h$ & Ellipticity of halo relative to light distribution,\\
 & ~~~~~$e_h/e_g$ \\
$f_{45}$ & Ellipticity of $\ds_{45}= 2f_{45}\ds_{iso}e_g\sin{(2\Dt)}$ \\
$f_{\gamma}$ & Ellipticity of shear before boosting; \\
   & $\ne f$ if boost has azimuthal dependence \\
$f_{int}$ & Ellipticity of distribution of \\
 & physically-associated ``sources'' \\
$f_{mag}$ & Ellipticity of magnification bias \\
 & $\delta N/N\propto\kappa\propto (1+2f_{mag}e_g\cos{(2\Dt)})$ \\
$f_{eff}$ & Effective ellipticity of $f_{int}$, $f_{mag}$ \\
   & taking into account dilution by isotropic \\
 & ~~~~~source distribution \\
\hline\hline
\end{tabular}
\end{table}

\section{Data}\label{S:data}

The data used for this paper come from the  Sloan Digital Sky
Survey, or SDSS \citep{2000AJ....120.1579Y}, an ongoing survey that will eventually image
approximately one quarter of the sky (10,000 square degrees).
Imaging data is taken in drift-scan mode in 5 filters, $ugriz$,
centred at 355, 469, 617, 748, and 893 nm respectively
(\citealt{1996AJ....111.1748F}, \citealt{2002AJ....123.2121S}) using a wide-field
CCD \citep{1998AJ....116.3040G} with photometric monitor
\citep{2001AJ....122.2129H}.   After the computation of an 
astrometric solution \citep{2003AJ....125.1559P}, the imaging data are
processed by a
sequence of pipelines, collectively called {\sc Photo}, that estimate
the PSF and sky brightness, identify objects, and measure their
properties. The software pipeline and photometric quality assessment
is described in
\cite{2004AN....325..583I}.  Bright galaxies and other interesting
objects are
selected for spectroscopy according to specific
criteria (\citealt{2001AJ....122.2267E};
\citealt{2002AJ....124.1810S}; \citealt{2002AJ....123.2945R}).   The
SDSS has had four major data releases: the Early Data Release or 
EDR \citep{2002AJ....123..485S}, DR1
\citep{2003AJ....126.2081A},  DR2
\citep{2004AJ....128..502A}, and  DR3
\citep{2005AJ....129.1755A}.

The lens and source catalogs are very similar to those described in M05, 
except for the inclusion of the fainter lenses with photometric
redshifts. Thus the description  
here will be brief, with the exception of the few differences from that 
work which will be described in more detail.

\subsection{Lens sample}\label{SS:lens}

While M05 used only $3\times 10^5$ spectroscopic galaxies ($r<17.77$) as lenses, this 
work uses a
larger sample of 2 million lenses as faint as $r=19$ (model magnitude) with
photometric redshifts from {\sc kphotoz}
v3\_2~\citep{2003AJ....125.2348B} in the range $0.02<z<0.5$.  There are
several reasons for this change.  First, in order to detect a
possibly small azimuthal variation in the tangential ellipticity, we
need a much larger sample of lenses.  Because of the higher number
density of lenses when we include these fainter lenses,
and because of the larger photometric area coverage, the use of a
fainter flux limit increases the size of our lens sample by a factor
of roughly eight.  Second, while a lensing analysis aimed at
determining the signal amplitude suffers from significant calibration
uncertainty when galaxies without spectroscopic redshifts are used as
lenses \citep{2005A&A...439..513K}, this work is primarily concerned with
the ratio of $f\ds$ to $\ds$, so errors in signal
calibration are irrelevant. 

There is one possible error in the detection of $f$ that could be
affected by the use of photometric redshifts.  The photometric
redshifts are used for each lens to determine $D_A(z)$ and
consequently the
transverse separation between lenses and sources.  If $f$ is a constant
value independent of radius for the projected mass
distribution, then errors in the computed transverse separation do not
matter, since they will affect $\ds$ and $f\ds$ in the same way.  However, if
$f$ decreases with radius, 
and errors in the photometric redshifts tend to go in one direction
(nonzero average bias), then the measured $f$
can be systematically affected.  If the photometric redshifts tend to
be biased high, 
then they will overestimate the value of $r$, and lead to an
overestimate of $f$; if they are biased low, they will underestimate
the value of $r$ and consequently of $f$.  Were it not for this
problem, we would make a larger lens sample by
going to fainter magnitudes, but  the bias and scatter in the
photometric redshifts at fainter magnitudes makes this impractical.
As shown in M05 using data from DEEP2, for $r<19$, the 
photometric redshifts are not noticably 
biased, and have scatter $\Delta z\sim 0.04$. 
Since $D_A(z)$ is not linear in redshift and $f$ is not linear in $r$,
in principle even with no bias and a scatter of $\Delta z\sim
0.04$, the errors in $r$ may bias our results, since we
use $r=\theta D_A(z)(1+z)$.  However, calculations indicate that this
potential bias is well below the statistical error.

The area covered by the full catalog is roughly 6200 square degrees.
Shape measurements were obtained for roughly 96 per cent  
of lenses passing the magnitude cut in this region, where many of the
failures were 
due to saturated centres or other problems for very bright galaxies.

The lenses were split into luminosity bins one model magnitude wide;
the notation for the bins here is the same as in M05, but we only use
the four brightest bins from that paper, for which the signal was detected with high
signal-to-noise.  The luminosities were computed using the photometric
redshift to get the distance modulus, and k-corrections are from {\sc
  kcorrect} v1\_11~\citep{2003AJ....125.2348B}.  As in M05, a 
luminosity evolution correction consistent with \cite{2003ApJ...592..819B} was
applied, shifting 
all $r$-band magnitudes by $+1.6(z-0.1)$.  Table~\ref{T:lenses} 
includes information about
the 
luminosity bins, including the numbers of
lenses and parameters of the redshift and magnitude distribution.  The
effective redshifts and luminosities are 
computed using the same weights as are used in the computation of the
lensing signal.  Information is shown for red and blue samples
separately, 
where colour separation will be described shortly.
\begin{table}
\caption{\label{T:lenses} For each luminosity bin, the number of lens galaxies,
 $\langle z\rangle$, and $\sigma(z)$ (a characteristic width,
  though the distribution is not Gaussian),
  the mean weighted redshift $z_{\rm eff}$, and the mean weighted luminosity
  $L_{\rm eff}$ relative to $L_*$.  The mean weighted
  redshifts differ for the different source samples, and the results
  shown are the average of those values.}
\begin{tabular}{ccllll}
\hline\hline
Sample, $\mr$ range & $N_{gal}$  & $\langle z\rangle$  &
$\!\!\sigma(z)\!\!$ ~ & $z_{\rm eff}$ & $\!\!\!L_{\rm eff}/L_*\!\!\!$ \\
\hline
\multicolumn{6}{c}{Blue lenses} \\
L3, $[-20,-19)$ & $\!\!$215 808 & 0.12 & 0.04 & 0.12 & 0.5 \\
L4, $[-21,-20)$ & $\!\!$310 161 & 0.19 & 0.05 & 0.17 & 1.1 \\
L5, $[-22,-21)$ & $\!\!$145 136 & 0.27 & 0.09 & 0.23 & 2.5 \\
L6, $[-23,-22)$ & 38 683 & 0.37 & 0.11 & 0.27 & 6.3 \\
\multicolumn{6}{c}{Red lenses} \\
L3, $[-20,-19)$ & $\!\!$232 446 & 0.13 & 0.04 & 0.12 & 0.5 \\
L4, $[-21,-20)$ & $\!\!$538 364 & 0.18 & 0.05 & 0.17 & 1.2 \\
L5, $[-22,-21)$ & $\!\!$431 775 & 0.25 & 0.08 & 0.23 & 2.6 \\
L6, $[-23,-22)$ & $\!\!$107 883 & 0.34 & 0.10 & 0.26 & 5.9 \\
\hline\hline
\end{tabular}
\end{table}

In addition to the real lens catalog, our analysis also
requires catalogs of random lens positions.  These were created by
distributing the random lenses uniformly across the survey area, since
the photometric survey at $r<19$ is fairly uniform and the density of sources at
bright magnitudes does not vary significantly with seeing (at fainter
magnitudes, where star/galaxy separation is more difficult, the
density of sources does not show the expected increase with magnitude,
and shows significant variation with seeing). 

\begin{figure}
\includegraphics[width=3in,angle=0]{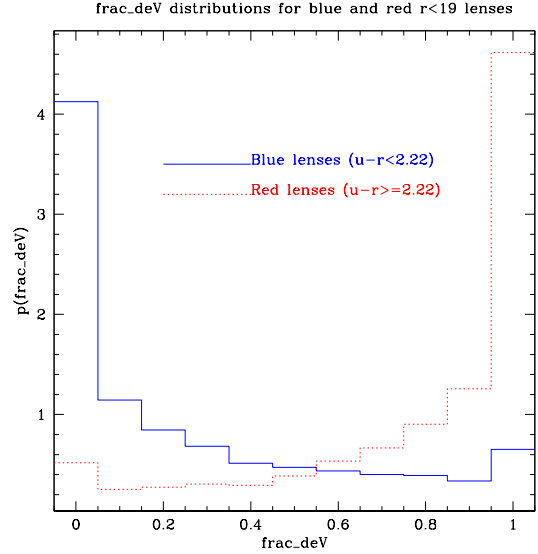}
\caption{\label{F:fracdev} The distribution of {\tt frac\_deV} values for
  red and blue samples of galaxies, respectively}
\end{figure}

It is also necessary to split the lens sample approximately into
elliptical and spiral samples.  For this purpose, we used the observed
$u-r$ (model) colour, with a division at 2.22 as in
\cite{2001AJ....122.1861S}.  As a check of this colour separator, we
used an independent method of classifying morphology, the {\tt frac\_deV}
parameter output by {\sc Photo} which, for each band, is determined
by fitting the profile to {\tt frac\_deV} times the best-fit deVaucouleurs
profile plus $1-${\tt frac\_deV} times the best-fit exponential profile, and
truncating the allowed range of values at 0 and 1.  This procedure is
done in each 
band; we use the $r$ band result due to its high signal to noise.
Thus, {\tt frac\_deV} is determined only via the 
profile shape in one band, not from the colour.  Fig.~\ref{F:fracdev} shows the
distribution of {\tt frac\_deV} values for the red versus for the blue
galaxies; as shown, the distribution for blue galaxies is strongly
peaked at zero and for red galaxies at one, as would be expected if
red colours and deVaucouleurs profiles are correlated, and if blue
colours and exponential profiles are correlated.  Hence we conclude that
our ``blue'' and ``red'' samples  include relatively pure samples
(at the $\sim 90$ per cent level) of spiral and elliptical galaxies
respectively.  

\subsection{Source sample}

The source catalog is the same as in M05.  In brief,
shape measurements were performed on all SDSS imaging data collected
through June 2004, using the re-Gaussianization technique described
in  \cite{2003MNRAS.343..459H}, yielding roughly $3\times 10^7$ source
galaxies.   
Star-galaxy separation was achieved using {\sc Photo} output
OBJC\_TYPE, and galaxies were required to meet the following criteria:
no flags from {\sc Photo} indicating problems with object measurement
(e.g., a saturated centre) can be set, 
$r$ band extinction less than
0.2, $r$ band extinction-corrected model magnitude  $r<21.8$, resolution
factor $R_2>1/3$ in both bands ($r$ and $i$) used for the shape
measurement to ensure high-quality shape measurements and to eliminate stars
that were mis-classified by OBJC\_TYPE as galaxies.  For the purpose
of the calculation of the shear signal, there are 3 
disjoint sets of sources: ``bright'' ($r<21$) sources, ``faint''
($21<r<21.8$) sources, and high-redshift ($0.4<z<0.65$) Luminous Red
Galaxies, or LRGs, which are explicitly excluded from the other two samples.  
The shear calibration was found in M05 to be accurate to
within $[-7, +12]$ per cent ($2\sigma$ level) for $r<21$ sources,
$[-10, +18]$ per cent for $r>21$ sources, and $[-8, 19]$ per cent for
high-redshift LRGs.  Redshift distributions were determined for faint
($r>21$) sources using DEEP2 spectroscopy
(\citealt{2003SPIE.4834..161D}, \citealt{2003ApJ...599..997M},
\citealt{2004astro.ph..8344D}, \citealt{2004ApJ...609..525C}), whereas
redshift estimates for individual galaxies were obtained for the other
two source samples.  
For the bright ($r<21$) sources, photometric redshifts from {\sc kphotoz}
v3\_2~\citep{2003AJ....125.2348B} with error distributions determined
using DEEP2 
spectroscopy in M05 were used; LRG 
photometric redshifts with error distributions from
\cite{2005MNRAS.359..237P} were used for the  LRG sample. 

\subsection{Systematics}\label{SS:systematics}

In M05, a large number of systematic errors in the weak lensing signal
were analyzed to assess their significance in the SDSS data.  For
this work, many of the calibration uncertainties considered there
(e.g., shear
calibration bias, redshift distribution systematics, and stellar
contamination) are irrelevant due to our interest in the ratio of
$f\ds$ to $\ds$, both of which have the same calibration biases.
However, there are other, more subtle systematics that are important here,
that we describe in the subsections that follow.

\subsubsection{Systematic shear}\label{SSS:sysshear}

One possible contaminant of the $f$ measurement is systematic shear.  As discussed in
M05, any average smearing of the images along the scan
direction will cause a slight alignment of the lens and source
ellipticities.  This systematic thus raises (lowers) the 
shear signal on large scales by an additive factor if the PSF correction scheme over-
(under-)estimates the correction to the shears due to this smearing. A
constant systematic shear
only changes \ds{} on large scales, because for an isotropic
distribution of sources around a lens, all contributions from the
systematic shear to the tangential ellipticities cancel out;
consequently, it is only important for lenses with an anisotropic
distribution of sources.  Fortunately, as described there, we can correct for
the effects of systematic shear in the lensing signal  using random
lens catalogs to compute the signal due to
systematic shear, so that we can subtract it off from the real signal.

Unfortunately, for the  $f\ds$ measurement, this procedure
is insufficient.  The alignment of lens and source ellipticities
due to systematic shear will lower $f\ds$ (since $\gamma_{t,sys}<0$ where
$\cos{(2\Delta\theta)}>0$, and vice versa) thus lowering our estimate
of $f$.  This signal cannot be calculated analytically and removed
without a very well-understood model for the size of the
systematic shear and its variation on the sky (which we lack).  However, there is one
way to measure it: the systematic shear contributes
equally to $f\ds$ and to $\fdsrot$ (since rotating the lens and
source ellipticities by the same angle does not change the fact that
they are aligned).  Since the spurious shear contributions to
$\fdsrot$ and $f\ds$ are equal,  we can subtract  $\fdsrot$ from
$f\ds$ to measure the uncontaminated value of 
$(f-f_{45})\ds_{iso}$, for which we presented model predictions in \S\ref{SS:elliptheory}.
As theoretical justification for the subtraction to eliminate systematics, we note
that the shear correlation functions are related via
\be
\xi_{++}(\theta) \pm \xi_{xx}(\theta) = \frac{1}{2\pi} \int \ell d\ell \left[
  P_E(\ell) \pm P_B(\ell) \right]
\begin{cases}
J_0 (\ell\theta) & \\
J_4 (\ell\theta). & 
\end{cases}
\ee
Here, $\xi_{++}$ and $\xi_{xx}$, the correlation functions of
tangential and 45-degree rotated shear, are related via multiplicative factors
to the measured $f\ds$ and $f_{45}\ds_{45}$ (respectively) resulting
from systematic shear, $P_E$ is
the power spectrum of the $E$-mode contribution to systematic shear, and
$P_B$ is the power spectrum of the $B$-mode systematic shear.
On small scales ($\ell\theta\ll 1$, so $J_0\rightarrow 1$ and
$J_4\rightarrow 0$), $\xi_{++}=\xi_{xx}$, i.e. the contribution of a
constant systematic shear to $f\ds$ and $\fdsrot$ is identical.  As we
will see, the signal from the systematic shear is indeed consistent
with being associated with scales larger than those we use
for the halo ellipticity measurement, justifying our scheme for eliminating
this systematic.

\subsubsection{Anisotropic source number density}
\label{sss:aniso}

In this paper, we are primarily concerned with transverse separations
from 20 \hkpc{} 
(the minimum scale at which signal is measured) to several
hundred \hkpc, around the virial radius. 
For our measurement, we use a boost
factor $B(r)$ (to account for dilution of the signal due to inclusion
of non-lensed galaxies in the source sample) that is azimuthally
averaged, and therefore does not include variations of the observed
galaxy number
density with $\Delta\theta$.   We consider here four
effects that may cause azimuthal variation of the number density of
galaxies around lenses, and their implications for our measurement of
halo ellipticity using an azimuthally-averaged $B(r)$: anisotropic
magnification bias, sky subtraction errors, density-shape intrinsic
alignments, and the inclusion of foregrounds in the source sample.

The first such concern that can cause 
azimuthal variation of the source galaxy density is magnification bias, the
effect of which is determined 
for each source sample as described in section 4.7.3 of M05.  The
magnification bias on average affects the amplitude of both \ds{}
and $f\ds$ by an overall factor (since we boost by $1+\xi_{ls}$
determined using random catalogs, and therefore we are assuming that
some of the real sources visible due solely to magnification bias are
not lensed, so we overestimate the 
average signal).  However, since magnification bias causes $\delta N/N
\propto \kappa$, and $\kappa$ varies azimuthally in our model, we must
ask whether our use of
azimuthally-averaged $B(r)$ biases the measurement of $f_h$.  If
the ellipticities of light and dark matter are aligned, then the 
magnification bias anisotropy will increase the source density
along the lens major axis more than along the minor axis.  This effect
can be modeled as an isotropic background of sources that
would be seen in the absence of lensing, plus an isotropic
distribution of physically 
associated sources, which together give us a boost factor
$B_{iso}(r)$.  Magnification bias leads to some additional boost factor, 
$B_{mag}(r)[1+2f_{mag}e_g\cos{(2\Dt)}]$.  However, in reality these
extra sources along the major axis are truly lensed, so there is no
need for our 
boost factor to have azimuthal variation to account for the azimuthal
variation in source number density.  Therefore, in the presence of
anisotropic magnification bias, our use of the angle-averaged
$B(r) = B_{iso}(r)+B_{mag}(r)$ will overestimate the lensing signal,
biasing both \ds{} and $f\ds$, but not $f_h$.  

\begin{figure}
\includegraphics[width=3in,angle=0]{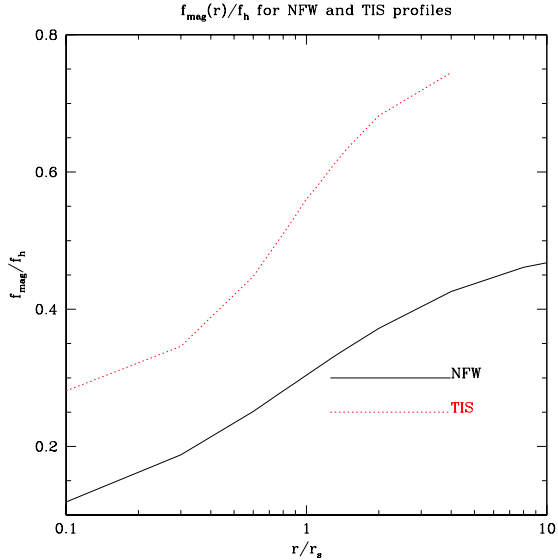}
\caption{\label{F:fmag} $f_{mag}(r)$ for the NFW and TIS models as a
  function of of transverse separation, in units of scale radii.}
\end{figure}

While anisotropy of the source distribution around lenses due to
magnification bias does not
bias the measurement of $f$ using an azimuthally-averaged boost
factor, it does provide an interesting test of 
our results if we can detect it, because the values of $f$ and
$f_{mag}$ are related to $f_h$ in a consistent way by the halo density
profile $\rho(r)$.  As shown in Eq.~\ref{E:kappamodel}, for a
purely power-law profile, $f_{mag}=f_h\alpha/4$.   Fig.~\ref{F:fmag}
shows $f_{mag}$ as  
a function of scale for NFW and TIS profiles.  When
studying our 
results in \S\ref{S:results} we will return to the issue of detecting
anisotropy of magnification bias.

The second effect we discuss that may cause number density anisotropy
  is small-scale errors due to {\sc Photo}, the sky subtraction
systematic described in section 6.3.7 of M05.  In short, a
problem with determination of the sky level near bright ($r<\sim
17.5$) lenses leads to variation of galaxy density within about  
1' due to errors in the sky flux scattering galaxies into or out of
  our flux and apparent size cuts.
Furthermore, model fits to determine fluxes and
shapes are influenced in a difficult to
quantify way.  Consequently, we cannot  assume an error in the
amplitude of \ds{} due to this effect that would cancel out of the ratio
$(f\ds)/\ds$, we must ask whether it will induce
alignments of densities on small scales that would change $f\ds$ but
  not \ds.  Furthermore, we must consider that this effect 
  depends on the angle from the 
lens major axis, since at fixed transverse separations, the light on
the major axis is brighter than on the minor axis. Fortunately, the
analysis is  simplified by the fact that in our $r<19$ lens sample
used for this work, only about 17 per cent are at $r<17.5$ where this
effect is a problem.  

We may naively
expect that if the sky systematic acts only affects number counts in
our catalog, then they will be changed more on the
major axis than on the minor axis.  However, this systematic
affects both lensed and physically-associated galaxies in the source
sample, so dilution due to physically-associated sources should not
have azimuthal dependence.  Consequently, our use of
azimuthally-averaged $B(r)$ is 
indeed appropriate.

The third effect we consider is the effect of intrinsic alignments.
There are two intrinsic alignment effects that must be taken into
account; the first is a correlation between the shapes of
physically-associated galaxies, or the ``shape-shape''
correlation; the second effect is a correlation between the shapes of
galaxies with the local density field, or the ``density-shape''
correlation. In this section we focus on the ``density-shape''
correlation, since it modulates the observed galaxy density around the
lenses, and defer discussion
of the ``shape-shape'' correlation for the next section.

While we attempt to eliminate physically-associated sources from our
calculation using photometric redshifts for the $r<21$ sample, the fact that our boost
factors are greater than one 
indicates that we are not fully successful.  If the distribution of
physically-associated sources around lenses is not isotropic, but is
instead aligned with the lens major axis, then the number density of
``sources'' on the sky may also appear to have such an alignment due to
contamination.  Such an effect is caused by an alignment between
the lens galaxy shape and the local density field.  This anisotropy of
the satellite  
distribution has actually been detected before, albeit with
conflicting amplitude and sign.  The initial
detection by \cite{1969ArAst...5..305H}, which gave rise to the name
``the Holmberg effect,'' was of an alignment of satellite galaxies along
the
minor axis of disk galaxies to transverse separations of 50 kpc.
Since then, several studies have shown a similar effect:
\cite{1997ApJ...478L..53Z} found alignment of satellites along the
minor axis of
spiral galaxies for separations of 300-500 kpc, and
\cite{2004MNRAS.348.1236S} found the same using a
mixed sample of satellites in the 2-degree Field survey (2dF) out to
500 kpc.  Other 
studies have shown the opposite; for
example, 
\cite{2005ApJ...628L.101B}, using a sample of satellites in
the SDSS to projected
separations of 400 \hkpc{}, found an alignment along the major axis;
\cite{2006MNRAS.367..611M} found a larger-scale alignment ($0.3$--$60$ $h^{-1}$Mpc) along
the major axis for $L > L_*$ spectroscopic galaxies in the SDSS. 
And finally, other works (\citealt{1975AJ.....80..477H} and
\citealt{1982MNRAS.198..605M}) have found isotropic satellite
distributions. These results were found using small primary-satellite
systems; results with clusters are more definitive, indicating a
tendency for the bright cluster galaxy (BCG) to be elongated along the
cluster major axis, and hence for it to have satellites concentrated
along $\Delta\theta=0,\pi$ and a deficit at $\pi/2$ and $3\pi/2$,
a manifestation of the Binggeli effect
\citep{1982A&A...107..338B,1999ApJ...519...22F,2000ApJ...543L..27W,2002ASPC..268..395K}.
An alignment of physically-associated galaxies with the lens major or
minor axis poses a problem for our measurement of $f$:
if there is a larger (smaller) number of physically associated ``sources'' along the major axis, our
azimuthally-averaged boost factor underestimates
(overestimates) the signal on the major axis and vice versa for the
minor axis, thereby underestimating (overestimating) $f$.  We attempt
to measure this effect using a test to be described at the end of this section.

The final effect we consider is the inclusion of galaxies in the
source sample that are actually
between us and the ``lens,'' and that lens the ``lens'' (foreground contamination). Because this
effect causes the ``lenses'' to have tangential ellipticity relative
to the ``sources,'' it may increase number counts of
``sources'' along the ``lens'' minor axis, and decreased number counts
along the major axis.  Because these foregrounds are not lensed, our
boost factor needs to have azimuthal dependence to account for them;
since it does not, we may overestimate $f$.  (They also lead to
overall suppression of the signal which is unimportant for our $f$
measurement.)  Due to our cuts, we do 
not expect there to be large numbers of foregrounds in our sample, so
this effect should not be a major problem, but still test our
assumption using the tests described below.

To determine which, if any, of these four possible causes of
azimuthally-dependent variations in source density actually occurs, we
use the following test.  For each lens sample and the
sources with photometric redshifts, we divided the lens-source pairs
into three samples: the ``foreground'' sample with $z_s <
z_l-\epsilon$ ($\epsilon = 0.05$, 0.1), the ``physically associated'' sample with $z_l-\epsilon
\le z_s \le z_l+\epsilon$, and the ``source''
sample with $z_s > z_l+\epsilon$.  Due to photometric
redshift errors, these divisions are imperfect.  For each of these
three samples and lens
luminosity/colour samples, we compute the
angle-averaged boost-factor $B(r)=N_{LS}/N_{RS}$, where
$N_{LS}$ is the number of pairs of real lenses and real sources, and $N_{RS}$
is the number of pairs of random lenses and real sources 
(with weight equal to 
1 for each galaxy), and we also compute the sum over pairs $\xi_{\Dt}(r) = \sum_{LS}
e_g \cos{(2\Dt)}/N_{RS}$.  

If we assume that the galaxy density has
some isotropic distribution $N_{iso}(r)$ plus a varying component
$N_{\Dt}(r)[1+2f_{sys}e_g \cos{(2\Dt)}]$ due to one of the
aforementioned effects, then we will have
\bea
B(r)&=[N_{iso}(r)+N_{\Dt}(r)]/N_{RS}(r) \\
\xi_{\Dt}(r) &= 2N_{\Dt}(r)f_{sys}e_{rms}^2/N_{RS}(r) \notag
\eea
(the factor of two comes from
our use of the full ellipticity $e_g^2$, the sum of the
squares of the two components, for which
$\langle e_g^2\rangle = 2e_{rms}^2$).  Comparing $\xi_{\Dt}(r)$ with $B(r)$ allows us to
determine the effective $f_{eff}(r)$ of the distribution of sources,
where $f_{eff}(r) = f_{sys} N_{\Dt}(r)/(N_{\Dt}(r)+N_{iso}(r))$ (if
the anisotropic component of the number density is much smaller than 
the total angle-averaged number density, the effect of the anisotropy
will be correspondingly smaller).  For the two sources of
anisotropy $N_{\Dt}(r)$ for which our use of azimuthally-averaged
boost factor is incorrect
(intrinsic alignments and foregrounds), we will have underestimated
$f$ and therefore $f-f_{45}$ by the $f_{eff}$ corresponding to that
effect, to lowest order in $e$.  This underestimate can be understood
considering that $\ds\propto
(1+2fe_g\cos{(2\Dt)})$, but if the boost factor has azimuthal variation,
then we really measure shear $\propto (1+2f_{\gamma}e_g\cos{(2\Dt)})$
such that when multiplied by the boost $\propto
(1+2f_{eff}e_g\cos{(2\Dt)})$, we get
the true $\ds$ ($f_{eff}+f_{\gamma}=f$).  When we neglect the $f_{eff}$ of the boost, we therefore
underestimate $f$ by this additive factor.  However, when we measure
$f_{45}\ds_{45} \propto f_{45}e_g\cos{(2\Dt)}$, to lowest order in
$e_g$ there is
no correction to account for our use of the azimuthally-averaged boost.  

We expect a variation in number density due to
magnification bias will appear predominantly on small scales in the
``source'' sample (most strongly for bright lenses); due to intrinsic alignments, mainly in the
``physically associated'' sample but with uncertain scale dependence; due
to {\sc Photo} effects in all three source samples equally. but only
on small scales; and due to lensing by foregrounds predominantly in
the ``foreground'' sample.  Results of these tests will be presented in
\S\ref{SS:resultssys}. 

\subsubsection{Shape-shape intrinsic alignments}

As mentioned in \S\ref{sss:aniso},
physically-assocated lens-source pairs may potentially contaminate our
measurement due to any shape-shape intrinsic alignments.  If the
degree of galaxy alignment is independent of $\Delta\theta$, then the
shape-shape alignment manifests in the same way as a constant
systematic shear, and will be eliminated by the
$f\Delta\Sigma-f_{45}\Delta\Sigma_{45}$ subtraction.  However, any
azimuthal structure in the shape-shape intrinsic alignments would not
be removed via this subtraction, and will contaminate the
measurement.  To test for this problem, we calculate
$f\gamma-f_{45}\gamma_{45}$ for sources in the
``physically associated'' sample (we measure this instead of
$f\ds-f_{45}\ds_{45}$ because $\Sigma_c^{-1}$ is zero for $z_s \le
z_l$).   Results of this test are presented in section~\ref{SS:shapeshape}.

\subsubsection{{\sc Photo}-related shape measurement biases}

In principle, the sky brightness systematic may
bias the shapes of sources in a way that would affect the
measurement of $f\Delta\Sigma-f_{45}\Delta\Sigma_{45}$.  This might
occur if the sources  align with the gradient of
the sky brightness near the lenses.

To test for this contamination, we measure
$f\gamma-f_{45}\gamma_{45}$ using sources in the ``foreground'' sample
(with  $z_s < z_l-0.1$).  This quantity should not have significant
contamination from actual lensing or from azimuthal 
structure in intrinsic alignments, and hence should be dominated by
software-related systematics.  Results of this test are presented
in section~\ref{SS:photobias}.

\section{Method of analysis}\label{S:analysis}

The first step for the halo ellipticity calculation is to compute
$\ds$, $f\ds$, and $\fdsrot$ as in Eqs.~\ref{E:sums} and~ref{E:sums45}
and the accompanying text.  As described in \S\ref{SSS:sysshear},
$\fdsrot$ will allow us to identify systematic shear
contributions to $f\ds$.  Covariance matrices for these quantities are
computed via bootstrap resampling with 150
subregions, similar to the method used in M05.  Once we compute these
quantities as a function of comoving pair separation $r$, 
there are several ways to proceed with the analysis.

The first method used is a non-parametric determination of $f-f_{45}$
averaged over radius.  We start with
the signal \ds{} for the desired range of radii, and $f\ds-\fdsrot$ with errors determined from those
quantities added in quadrature (they are uncorrelated).  We then consider the problem of two Gaussian variables
$y$ and $x$ that are related via $m=y/x$ (here $y$ represents
$f\ds-\fdsrot$, $x$ represents $\ds$, and $m$ is our
desired quantity, $f-f_{45}$).  In our case, we have multiple estimators
$\hat{y}_i$ and $\hat{x}_i$ (the values at each radial bin) and
would like to  combine them while taking into account
non-Gaussianity in the ratio $\hat{y}/\hat{x}$ (\citealt{bliss1},
\citealt{bliss2}, \citealt{fieller}).  For each measurement
$i$, the quantity $\hat{y}_i-m\hat{x}_i$ is a random Gaussian variable
drawn from an $N(0,\sigma_{y_i}^2+m^2\sigma_{x_i}^2) \equiv N(0,w_i^{-1})$
distribution.  Consequently, the following summation over all measurements is
also a random Gaussian variable:
\be
\frac{\sum w_i (\hat{y}_i-m\hat{x}_i)}{\sum w_i} \sim 
N\left(0,\frac{1}{\sum w_i}\right)
\ee
where the distribution is taken at fixed $m$, and the weights $w_i$
depend on $m$.  We can then determine frequentist confidence
intervals\footnote{We are 
not constructing a Bayesian posterior region $P(m)$.} at the
$Z\sigma$ level by writing
\be\label{E:confint}
\frac{-Z}{\sqrt{\sum w_i}} < \frac{\sum
  w_i(\hat{y}_i-m\hat{x}_i)}{\sum w_i} < \frac{Z}{\sqrt{\sum{w_i}}}.
\ee
Since this expression depends on $m$, in order to calculate
the confidence intervals we make a grid in $m$ and interpolate to find
the value of $m$ for the desired 
confidence interval, e.g. $m(Z=0)$ corresponds to the average value,
$m(Z=\pm 1)$ give the 68 per cent confidence limits, and so
on.  The value of this approach is that it gives a radius-averaged
value of $f-f_{45}$ without requiring a model for the shape of the profile or
of $f-f_{45}$, that it easily allows us to combine measurements of $f-f_{45}$ with
different source samples for the same lenses, and that it takes into
account non-Gaussianity of the error distributions of the ratios $\hat{y}_i/\hat{x}_i$. Once we have measured
$\langle f-f_{45}\rangle_r$ nonparametrically, we relate the result to
the predictions for different profiles to extract the value of interest, $f_h$.

The second method used here is to fit the signal to a power-law
profile using the predicted relationships in
Eq.~\ref{E:gammapowerlaw}.  That is, we write
\begin{align}\label{E:jointfit}
\ds(r) &= Ar^{-\alpha} \\
f\ds(r)-\fdsrot(r) &= Af_h\left[\frac{(\alpha-2)^2}{4\alpha}\right]r^{-\alpha}\notag
\end{align}
and fit jointly for $A$, $\alpha$, and $f_h$.  Unlike the previous
method, it does not take into account non-Gaussianity in the error
distribution of $f_h$, which we will see is important (and we cannot
use this method to measure $f_h$ with the NFW or TIS density profiles).

Finally, we repeat this analysis on large scales (800 \hkpc{} to 2
\hmpc) with all samples to check for alignment of the light
distributions of spirals and ellipticals with local LSS.

\section{Results}\label{S:results}

\subsection{Number density systematics}\label{SS:resultssys}

Here we present
results related to the azimuthally-dependent boost factor tests
described in \S\ref{SS:systematics}.  As described there, for
``foreground,'' ``physically associated,'' and ``source'' samples
(determined using photometric redshifts), we compute the effective
$f_{eff}(r)$ with each of the lens samples; this
computation allows us to determine whether magnification bias, {\sc Photo}
effects, intrinsic alignments, and lensing by intervening foregrounds
may be significant. 

\begin{figure}
\includegraphics[width=3.2in,angle=0]{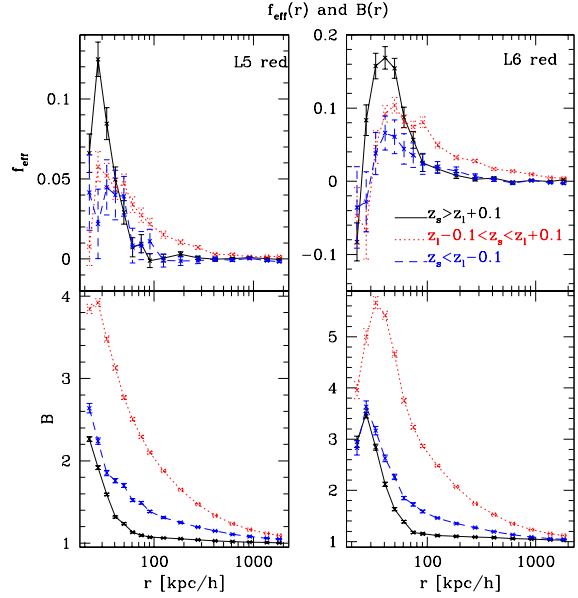}
\caption{\label{F:fboost} The measured $f_{eff}(r)$ and $B(r)$ for red
lenses, L5 and L6, with the three different sets of sources:
foregrounds, physically associated, and sources.  Note the different
scales on the vertical axes for L5 and L6.}
\end{figure}

For blue lenses of all luminosities, and for L3 and L4 red lenses,
$f_{eff}(r)$ was found to be consistent with zero on all scales, as
will be discussed further.  For
red L5 and L6 lenses, this was not the case; a plot of $f_{eff}(r)$ and $B(r)$
for these two lens samples is shown in Fig.~\ref{F:fboost} with
the three source samples.  

There are several important features in this plot.  First, we
consider the bottom two plots, $B(r)$ for L5 and L6.  The fraction of
physically associated sources in a given sample is equal to
$(B-1)/B$.  So, we see that our
photometric redshift cuts are reasonably efficient at isolating a sample of
physically associated sources;
at 100 \hkpc, in L6, nearly $2/3$ of the ``physically associated'' sample
is actually physically associated, compared to about 15 per cent of the
``source'' sample and 35 per cent of the ``foreground'' sample.  
The fact that $B(r)$ for
the physically associated sample actually increases with radius for
$20 < r < 30$ \hkpc{} in L5 and $20 < r < 40$ \hkpc{} in L6 indicates that
these ranges of radii are affected either by the sky subtraction
problem, or by the actual light from the lens galaxy; in any case, due to the
possibility of photometric errors in these regions, we exclude them
from the analysis in the rest of the paper.

\begin{figure}
\includegraphics[width=3in,angle=0]{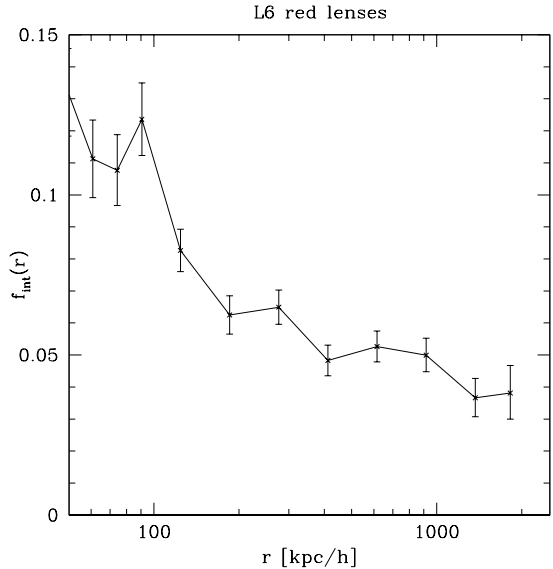}
\caption{\label{F:fint} The measured $f_{int}(r)$, including non-Gaussian errorbars, for  L6 red
lenses derived from the results for physically associated
sources}
\end{figure}

Next, we consider the top plots in Fig.~\ref{F:fboost}, $f_{eff}(r)$ for L5 and L6 red
lenses.  For L6, we can immediately see that for $100<r<2000$ \hkpc,
$f_{eff}(r)$ is positive for the ``physically associated'' sample in a way
that is highly statistically significant, but not for the other
samples.  This fact points to density-shape intrinsic alignments as the
cause for this finding.  Apparently, there is a tendency for satellite
galaxies in these ranges of radii to cluster along the major axis of
the ``lens'' galaxy.  The results shown here include everything from
small groups to large clusters, so what we see is an average of this
effect over systems of all sizes.  Another work (in preparation) will further explore the
luminosity and environment-dependence of this effect in an attempt to
understand the cause.  This plot shows $f_{eff}$; the actual value
of $f_{int}$ associated with intrinsic alignments (assuming that the number density on the sky about lenses is
isotropic with the exception of this effect) is determined via
$f_{eff}(r) = (B(r)-1)f_{int}(r)/B(r)$.  Fig.~\ref{F:fint}
shows $f_{int}(r)$ for this sample.  

We could also attempt to derive
similar results from the foreground or source sample (at least for
$r\sim 100$ \hkpc, where there is a slight signal) for the sake of comparison.
Since $B(r)$ is much lower, $f_{eff}(r)$ due to intrinsic alignments
is much lower for these samples.  When the results are compared for
these radii, the resulting $f_{int}$ is consistent with that shown in
Fig.~\ref{F:fint} at the 
$1\sigma$ level (we may not expect them to {\it a priori} be
identical, since the samples are different: the physically associated 
sample includes those with correct photometric redshifts, whereas the
foreground and source sample includes physically associated sources
with incorrect photometric redshifts, and since they lie in
separate regions of colour space, they may have different properties with
respect to intrinsic alignments).  

For L6, it is clear that on scales larger than 100 \hkpc, the
anisotropy of the number density distribution around lenses is due to
density-shape intrinsic alignments.  Now, we consider scales $r<40$ \hkpc; we
already know that the sky subtraction effect or the light of these
lenses is causing a loss of number density (as in the $B(r)$ plot), and the rapid decrease in $f_{eff}(r)$ with
decreasing $r$ on these scales indicates that the suppression of number
density is stronger along the major axis of the lens light
distribution, as may be expected.  This explanation is supported by
the fact that on these scales $f_{eff}(r)$ has roughly the same pattern for all three
source samples, and therefore the cause cannot be a physical effect
on only foregrounds, physically associated galaxies, or lensed
sources.  This finding supports our decision to not use these scales
when attempting to measure halo ellipticity.

Finally, we consider $40 < r < 100$ \hkpc.  For this range of
radii, $f_{eff}(r)$ changes with $r$ in similar ways for the foreground
and physically associated samples; this suggests that it is due to
physically associated sources in both, with the amplitude difference
being due to the different fraction of physically associated sources in
the two samples. 
However, as $r$ decreases in this range, $f_{eff}(r)$ increases much
faster for the lensed sources; this fact suggests that the increase is
due to anisotropy of the magnification bias causing an increase in
source number density preferentially along the lens major axis.
Unfortunately, the exact $f_{eff}$ for this effect is difficult to
extract from the plot due to the unknown radial dependence of the
$f_{eff}$ due to intrinsic alignments.  We could also attempt to
predict it, using the assumed value of $f_{mag}(r)$ for a given
profile, with the appropriate $\delta N/N$, to get 
\be
f_{eff,mag}(r) = f_{mag}(r) \frac{\delta N/N}{B(r)}
\ee
(The division by $B(r)$ is because we need the ratio of the additional
lensed sources to the total number of pairs, not to the total number of
lensed pairs which is given by $\delta N/N$.)  The difficulty is
that it is hard to untangle the effects of
$f_{int}$ and $f_{mag}$ in a model-independent way, which means that
(1) we cannot easily use the
measured $f_{mag}$ as a check on the results for $f$, and (2) we
cannot easily correct the measured $f$ for the effects of
$f_{int}$ (since it, but not $f_{mag}$, contaminates the measurement
of $f_h$).

We can consider several approaches to this problem, but the simplest
is to assume that the intrinsic
alignments in all three source samples are similar (i.e. there is no 
colour
or magnitude dependence), and therefore to get $f_{eff,int}(r)$ for
the lensed sources, we can
just use the implied $f_{int}(r)$ from the ``physically-associated''
sample in Fig.~\ref{F:fint} with
$B(r)$ to get $f_{eff,int}(r)$ for any source sample. Comparing
this prediction against the measured $f_{eff}(r)$ for a given source sample
allows us to untangle the effects of $f_{int}$ and $f_{mag}$.  

For L5, the situation is similar, though (1) the small-scale
suppression of number counts
only extends to 30 \hkpc, not 40 \hkpc, and (2) the detection of
ellipticity of the satellite galaxy distribution (density-shape
alignment) implied by the
nonzero $f_{eff}(r)$ for the physically associated sample is not as
large as for L6, though still statistically significant for $r< 400$ \hkpc.
There is a hint of magnification bias anisotropy on
scales up to 70 \hkpc{} for this luminosity bin.

The results shown in Fig.~\ref{F:fboost} imply that we do not observe significant
lensing by intervening foregrounds that may have biased our $f_h$
measurement.  This  may be because the SDSS data
are  rather shallow.

We now ask how these findings will affect the results with
fainter sources as well.  There are three effects that we have
observed: small-scale effects due to sky subtraction or the lens light 
distribution causing a loss of sources preferentially along the major
axis; effects on all scales due to contamination by physically
associated sources that tend to align along the lens major axis;
and anisotropic magnification bias for $r<100$ \hkpc{} (L6 red lenses), which
increases source number density preferentially along the lens major axis.  (Since foreground
contamination is not a problem for the $r<21$ sources, at the lowest
redshifts, they are certainly not a problem for the fainter, higher redshift source
samples.)  We consider all three effects separately.

First, we expect that the fainter sources will be affected by the
small-scale light problems similarly to the $r<21$ sources used in
this section.  Consequently, for these sources we also avoid $r<30$
\hkpc{} (L5 red lenses) and $r<40$ \hkpc{} (L6 red lenses), but use
all scales above $20$ \hkpc{} for other lens samples.

Second, the $f_{eff}(r)$ due to physically associated sources is the
product of two factors, $f_{int}(r)$ and $(B(r)-1)/B(r)$.  For the
fainter sources, $B(r)$ is about 30 per cent higher for the fainter
sources (due to the lack of photometric redshift information) than for
the brighter ones for $r>60$ \hkpc; below that separation, the
difference is not as large.  For small boosts as in all
luminosity bins for $r>100$ \hkpc{} and even for smaller scales for L3
and L4, this translates roughly to a doubling of the $(B(r)-1)/B(r)$
factor used to determine the $f_{eff}$.  However, it is not clear that
the $f_{int}(r)$ for these sources should be the same as for the
brighter sources.  To determine whether this is a safe
assumption, we repeated the tests shown above with $r<20.3$ and
$20.3<r<21$ sources separately to check for evolution of $f_{int}$
with apparent magnitude.  In both L5 and L6, while $f_{eff}(r)$ and
$B(r)$ were different for the $r<20.3$ and $r>20.3$ sources, the
implied $f_{int}(r)$ values were consistent with each
other at the $1\sigma$ level for all ranges of radii and all luminosity samples that we use
for this analysis.
We thus use the same $f_{int}(r)$ for $r>21$ sources as for $r<21$ sources; if this is not the case,
we will see that the correction we apply is quite small compared to
our statistical errorbars. 
For LRG sources, the lens and source samples are essentially uncorrelated, so
contamination by physically associated sources is not a concern.

Finally, the magnification bias effect depends on two things: the
magnitude distribution of the source sample, and the $R_2$
distribution.  In M05, we derived that the predicted
$\delta N/N$ due to magnification bias is $1.9\kappa$, $0.7\kappa$,
and $3.1\kappa$ for $r<21$, $r>21$, and LRG sources in our catalog,
respectively.  $\kappa\propto\scinv$, which also varies with source
sample.

When we actually use the data, we must use the {\it
  weighted} (i.e., with weights used for lensing) $B(r)$ and
$f_{eff}(r)$ when doing the corrections.  The weighted $B(r)$ for the
data can be significantly lower than the unweighted, since the
$\Sigma_c^{-2}$ weighting weights higher-redshift sources more heavily, and
those sources are less likely to be physically associated with the lenses.

We want to use the
weighted $B(r)$ and the angular correlations to determine
$f_{eff,int}(r)$, the portion of the anisotropy in the source
distribution that is due to contamination by
anisotropically-distributed physically-associated pairs,
{\it excluding} the portion of $f_{eff}$ that is due to magnification
bias.  We then add $f_{eff,int}$ to the measured $f-f_{45}$ to find
the non-contaminated value.  The correction we calculate is well
below the $1\sigma$ errors, so too much importance should not
be placed on details of the correction scheme, which we outline
briefly here.  We assume that $f_{int}$ is independent of
source sample, and $f_{mag}$ for each sample must be
related by consistency relations due to its dependence on the
invariant $\Sigma$.  
We fit the $f_{int}$
value to an empirical model (linear in $\log{(r)}$, which has no
theoretical meaning) and extrapolate the values
of $f_{int}$ to lower radii to see what $f_{eff,int}$ they predict,
assuming that the residual $f_{eff}$ is due to magnification bias. 

The resulting $f_{eff,int}(r)$, 
the amount by which we underestimate the measured $f-f_{45}$ due to
our use of the azimuthally-averaged $B(r)$, is then 
averaged over the range of radii used for the measurement.  We
find that $f(r)-f_{45}(r)$ is underestimated by 0.01
for $r<21$ and $r>21$ sources with L5 red lenses, and by 0.02 for
$r<21$ and 0.03 for $r>21$ sources with L6 red lenses.  Intrinsic
alignments are not a significant contaminant for the LRG sources
because they are at higher redshifts than the lenses. 
All of the results in the following sections are automatically corrected for this
effect.

\subsection{Systematic shear}
\begin{figure}
\includegraphics[width=3in,angle=0]{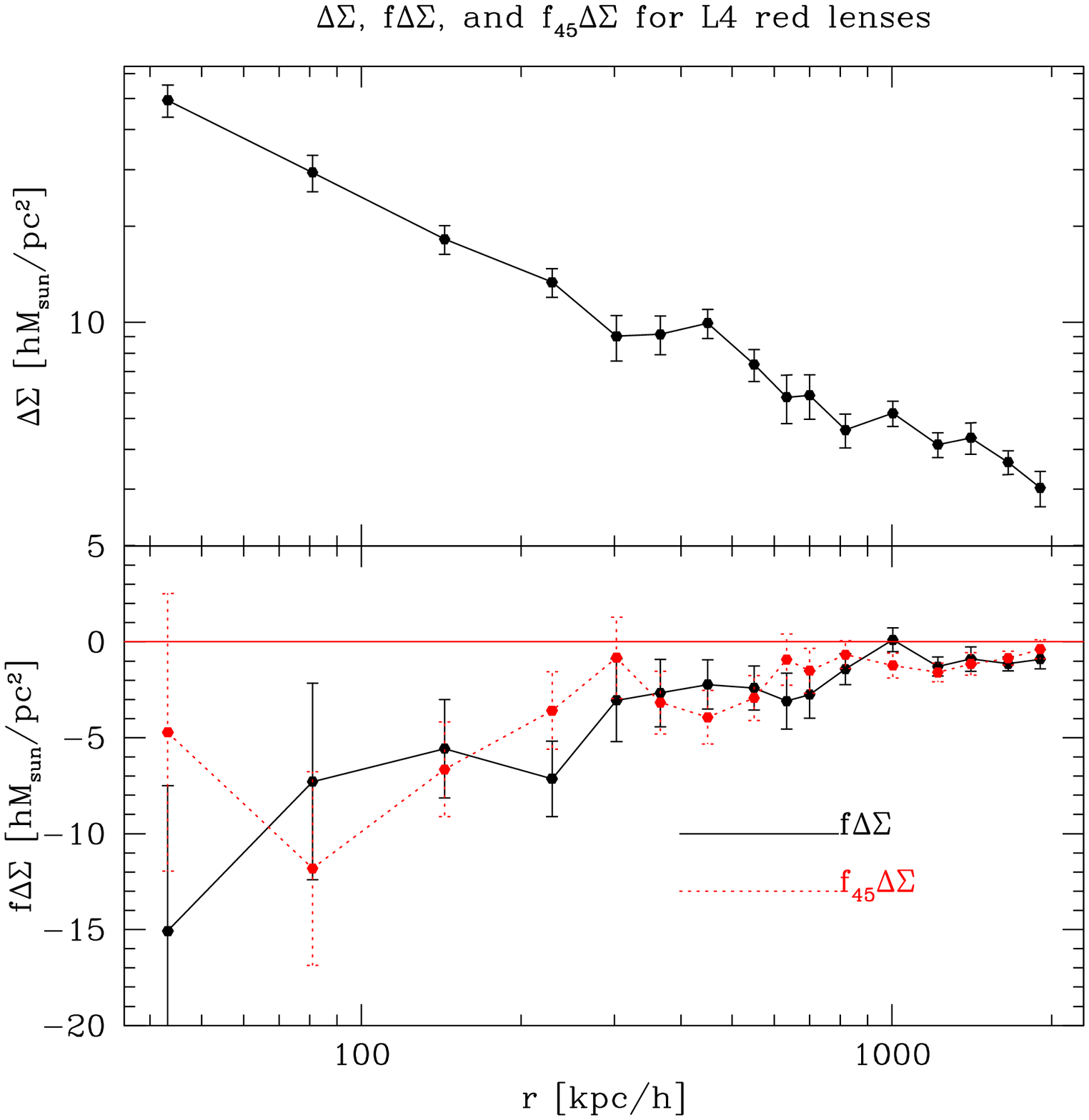}
\caption{\label{F:dsandfds} Plot of measured \ds{} (top), $f\ds$, and
  \fdsrot{} (bottom) for red L4 lenses with $r<21$ sources.}
\end{figure}

Here we present results for $\ds$, $f\ds$, and $\fdsrot$ for the 4
luminosity bins and 2 colour samples.  As an example of our findings,
Fig.~\ref{F:dsandfds} shows these three quantities for L4 red
lenses with $r<21$ sources.  As shown, $\fdsrot$ is clearly negative
at small transverse separations, and becomes closer to
zero with increasing radius.

We must then ask what could be causing this signal.  There are three
possibilities to consider.  First, we ask if cosmic shear could be
causing this result (i.e. lensing by LSS at redshifts lower than the
lens redshifts, causing real alignment of lens and source
ellipticities), so that the $e_{sys}$ that we are detecting is 
an actual effect rather than an artifact of some aspect of the shape
measurement.  We computed the expected cosmic shear signal for the
combinations of lens and source samples with the highest expected
cosmic shear signal, assuming the best fit
6-parameter $\Lambda$CDM cosmology from \citet{2005PhRvD..71j3515S},
the transfer function from \cite{1998ApJ...496..605E}, the growth
function from \citet{1992ARA&A..30..499C}, and the non-linear mapping
from \citet{1996MNRAS.280L..19P}.  The results in all cases indicate
that cosmic shear is more than 10 times too small to cause the
observed lens-source alignment.

Another possibility to consider is that the spurious ellipticity
$e_{sys}$ has been imparted to the lenses and sources in the course of
PSF correction.  The
PSF is determined in a given frame (10' by 14', or 1489 by 2048
pixels) using $\pm 2$ frames (5 frames total) to determine the Karhunen-Lo\`{e}ve (KL)
eigenfunctions, then using $\pm 1$ frame (3 total) to do a
second-order polynomial fit to the coefficients.  With 15--25 stars
per frame
used to determine the PSF \citep{2001ASPC..238..269L}, the average
separation between stars used to  determine the
PSF is $\sim 2.6'$.  There are a few ways
that incorrect PSF determination could cause this effect.   The first
is statistical: stars from $r\sim 16$ to $19$ are used to 
determine the PSF.  In our source catalog, the error on the
ellipticity for a galaxy at $r\sim 19$ is roughly 0.02.  If even half
of the stars used to determine the PSF are in 
the range $18<r<19$, the statistical error on the
PSF ellipticity is of the correct order to cause the effect we see.
The spurious ellipticity $e_{sys}$ due to this effect will of course
vary randomly from star to star, but since all stars are used to
determine the PSF variation across a frame, the statistical error due
to a single star may affect a region larger than the separation
between stars (since $\theta_{PSF}\sim 300$ $h^{-1}$kpc at the mean
effective redshift of the sample in Fig~\ref{F:dsandfds}).

 PSF systematics may also cause  this effect; the
PSF determination is accurate to within 2 per cent, so
systematics in the PSF determination  could give
rise to some ellipticity of 0.01--0.02.  These systematics include the
possibility that the spatial change in the PSF is not really
quadratic, or strange effects near the edges of the chip.

Now that 
we know that $\fdsrot$ and $f\ds$ are contaminated in similar ways on small
scales, 
our approach must be to
use $f\ds-\fdsrot$, and compare against model predictions
  for this quantity on small scales where $\theta \ll \theta_{corr}$, of order
  a few hundred \hkpc.

As a further test of the systematic shear subtraction technique, we
measured the correlation function of the systematic shear from the
data.  To do so, we used data from the southern galactic
regions, for which there were multiple ($>20$) observations of the same
area.  For $30$ pairs of runs in this region (selected randomly), we
isolated the set of objects with shape measurements in both runs.  For
those objects, we then found pairs of galaxies as a function of
angular separation.  In each run, we computed $e_t$ and $e_x$ for each
galaxy in the pair relative to the other, and constructed differences
$\Delta e_t$ and $\Delta e_x$ for each pair, where the $\Delta$
signifies the difference between the values of $e_t$ and $e_x$
in the two  runs.  Thus, $\Delta e_t$ and $\Delta e_x$ encode information
about systematic differences in the ellipticities of the same
galaxy from run to run.  For each pair, we then
used the values of $\Delta e_t$ and $\Delta e_x$ to construct
correlation functions
\begin{align}
\xi^{(sys)}_{++}(\theta) &= \langle \Delta e_{t,1} \Delta e_{t,2} \rangle \\
\xi^{(sys)}_{xx}(\theta) &= \langle \Delta e_{x,1} \Delta e_{x,2} \rangle \\
\xi^{(sys)}_{+x}(\theta) &= \langle \Delta e_{t,1} \Delta e_{x,2} \rangle 
\end{align}
where 1 and 2 denote galaxies in the pair.  

\begin{figure}
\includegraphics[width=3in,angle=0]{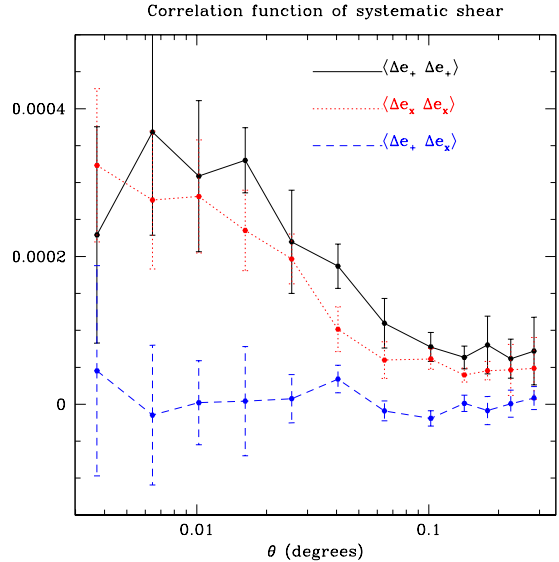}
\caption{\label{F:ellipcorrfun} Plot of measured correlation functions
  of the systematic shear as described in the text.}
\end{figure}

Fig.~\ref{F:ellipcorrfun}
shows the results of this computation, divided by two  to
approximate the result for the real data, because when
comparing the ellipticities across runs, the systematic shear in each
of them presumably adds coherently.   As shown, $\xi^{(sys)}_{++}$
and $\xi^{(sys)}_{xx}$ are consistent with each other, and
$\xi^{(sys)}_{+x}$ is consistent with zero, as expected for any
parity-preserving systematic such as spurious ellipticity along the
scan direction.  Furthermore,
the decline of the correlation functions with angular scale indicates
that we are correct in our understanding that this is a small-scale
systematic.  For reference, we note that in L3, the typical radial
range used for our calculations (20--280 \hkpc) corresponds to
$0.0055<\theta<0.08$ degrees, and in L6, 40--320 \hkpc{} corresponds to
$0.0040<\theta<0.03$ degrees.  We see from the figure that it is possible
that $\langle \Delta e_x \Delta e_x\rangle$ is lower than $\langle
\Delta e_+ \Delta e_+\rangle$ by roughly $5\times 10^{-5}$, and have verified
that the resulting changes in the computed values of $f_h$ are
significantly less
than half of the $1\sigma$ errors for all models, where our overestimate of
$\langle \Delta e_x \Delta e_x\rangle$ translates to an overestimate
of $f_h$.

\subsection{Shape-shape intrinsic alignments}\label{SS:shapeshape}

Here we present results of the calculation of $f\gamma-f_{45}\gamma_{45}$
for the ``physically associated'' source sample.  We then
compare the value of 
$f\gamma-f_{45}\gamma_{45}$ to the observed $\Delta\Sigma$ to see what
constraints we can place on the 
contamination of $f_h$ from azimuthal structure in shape-shape intrinsic
alignments, $f_h^{(II)}$.

The resulting $f\gamma-f_{45}\gamma_{45}$ is consistent with zero at
the $1\sigma$ level for all color and magnitude lens samples, on all scales.  This
result suggests that the contamination of $f_h$ must be consistent
with zero as well;
nonetheless, since the predicted values of $f-f_{45}$ for some
profiles are so small, we cannot formally place strong
constraints on the contamination. For example, for the red lenses,
averaged over luminosity, the constraint on $f_h^{(II)}$ is $0.3\pm
0.6$.  For the individual luminosity bins, again for the red lenses,
the resulting values of contamination are $f_h^{(II)}=0.0\pm 1.8$,
$0.35\pm 0.81$, $0.57\pm 1.0$, and $-0.7\pm 1.7$ for L3--L6
respectively.  

We note that a recent work \citep{2006MNRAS.367..611M}, indicated no
evidence of shape-shape intrinsic alignments on 0.3--60 $h^{-1}$Mpc
scales using a low-redshift
($z\sim 0.1$) sample of galaxies from the SDSS, despite the detection
of a statistically significant density-shape
alignment.  However, this measurement did not extend
below the virial radius, and involved pairs of bright galaxies rather
than one relatively bright and one significantly fainter, so we cannot
rely solely on it. 

\subsection{{\sc Photo}-related biases}\label{SS:photobias}

In this section, we present results of the measurement of
$f\gamma-f_{45}\gamma_{45}$ using the ``foreground'' sample selected using
photometric redshifts.  Any measurable $f\gamma-f_{45}\gamma_{45}$ should thus be due
to software-related biases, e.g. problems in determination of sky
brightness or deblending errors.

As for the previous section, we found that the measured
$f\gamma-f_{45}\gamma_{45}$ for this sample was completely consistent
with zero on all radii, for all lens samples.  However, when we compared it against the real
$\Delta\Sigma$ for the ``source'' sample, we found that we cannot
place very strong constraints on the contamination of $f_h$ due to
photometric errors, $f_h^{(ph)}$.  Table~\ref{T:photocontam} shows the constraints for
each color and luminosity sample, computed using the SIS profile predictions for
$f-f_{45}$ (results with an NFW or TIS model are qualitatively similar).
\begin{table}
\caption{\label{T:photocontam} Central values and 68 per cent CL
  limits on the contamination of $f_h$ due to software biases.}
\begin{tabular}{crr}
\hline\hline
Luminosity sample & $f_h^{(ph)}$ (blue) & $f_h^{(ph)}$ (red) \\
\hline
L3 & $3.7_{-5.5}^{+5.7}$ & $1.3\pm 3.5$\\
L4 & $-0.9_{-4.2}^{+4.1}$ & $-1.0\pm 1.1$ \\
L5 & $-0.6\pm 2.4$ & $-0.9\pm 0.9$ \\
L6 & $-0.4_{-2.7}^{+2.6}$ & $-0.4_{-0.9}^{+0.8}$ \\
All & $0.0\pm 1.8$ & $-0.7\pm 0.6$ \\
\hline\hline
\end{tabular}
\end{table}

\subsection{Halo ellipticities}
Having computed the signal, we  proceed with a basic
analysis: a non-parametric determination of $f-f_{45}$ averaged over
radial ranges for each lens
sample and source sample combination, as 
in \S\ref{S:analysis}.  
There are several considerations for
choosing radial ranges: first, we prefer scales small
relative to $\theta_{corr}$, so that the $f\ds-\fdsrot$
subtraction is justifiable; second, that it is not clear what signal
to expect on scales for which neighboring galaxies in
the same halo contribute significantly to the lensing signal, so we
restrict to radial ranges in which the majority of the weak
lensing signal is coming from the lens alone.  For L3--L4, we start the measurement from the minimum
pair separation used here, 20 \hkpc, and go as far as 245 \hkpc{} (L3),
and 300 \hkpc{} (L4, L5); for L5, we use 30-300 \hkpc, and for L6, we
use 40-300 \hkpc{} because of the small-scale systematics noted in \S\ref{SS:resultssys}.  The maximum
separations are all smaller than $\theta_{corr}$, so the
subtraction of \fdsrot{} accurately removes contamination due to
systematic shear.  Table~\ref{T:fnonpar} shows the average $f-f_{45}$
for each lens and source sample combination. 

\begin{table*}
\caption{\label{T:fnonpar} Results for $f-f_{45}$ averaged over radial
 ranges described in the text for each combination of lens and source
  samples.  All errors shown are 68 per cent confidence intervals}
\begin{tabular}{cllll}
\hline\hline
Lenses & $r<21$ sources & $r>21$ sources & LRGs & all \\
 & \multicolumn{4}{c}{$f-f_{45}$} \\
\hline
L3, blue & $-0.04\pm 0.30$ & $-0.42_{-0.48}^{+0.43}$ & $-2.1_{-3.0}^{+2.3}$ & $-0.29_{-0.27}^{+0.26}$ \\
L4, blue & $-0.64_{-0.48}^{+0.43}$ & $-0.23_{-0.36}^{+0.35}$ & $-0.28_{-0.70}^{+0.64}$ & $-0.36_{-0.26}^{+0.25}$ \\
L5, blue & $-0.01\pm 0.43$ & $-0.36_{-0.44}^{+0.41}$ & $-0.35_{-0.58}^{+0.54}$ & $-0.27\pm 0.28$ \\
L6, blue & $0.57_{-0.80}^{+1.1}$ & $0.77_{-2.5}^{+4.0}$ & $1.82_{-0.9}^{+1.5}$ & $1.0_{-0.9}^{+1.3}$ \\
\hline
L3, red & $-0.12\pm 0.21$ & $-0.51_{-0.21}^{+0.20}$ & $-0.25_{-0.24}^{+0.23}$ & $-0.33_{-0.13}^{+0.12}$ \\
L4, red & $-0.08\pm 0.12$ & $0.07\pm 0.11$ & $0.04\pm 0.15$ & $0.01\pm 0.07$ \\
L5, red & $-0.19\pm 0.13$ & $0.12\pm 0.12$ & $0.38\pm 0.16$ & $0.08\pm 0.08$ \\
L6, red & $0.04\pm 0.22$ & $0.34\pm 0.16$ & $0.41\pm 0.24$ & $0.29\pm 0.12$ \\
\hline\hline
\end{tabular}
\end{table*}

Several results in this table require comment. First, the $f-f_{45}$
values using $r<21$ sources appear more negative than
those for the other source samples, so we must check for consistency of the samples.  We do this
by finding, for source samples $i$, 
\be
\langle f\rangle_{\mbox{source}} = \frac{\sum_i f_i/\sigma_i^2}{\sum_i
1/\sigma_i^2}
\ee
then getting 
\be\label{E:chi2test}
\chi^2 = \sum_i \left( \frac{(f_i-\langle f\rangle_{\mbox{source}})^2}{\sigma_i^2}\right)
\ee
which should follow a $\chi^2$ distribution with degrees of
freedom equal to one less than the number of samples being compared.
We indeed find that the results with the different 
samples for each lens sample are statistically consistent.

While the $\chi^2$ test is the most general one that is sensitive to
all kinds of discrepancies, the fact that 5 out of the 8 values of
$f_h$ for the $r<21$ sources are lower than for the $r>21$ sources
suggests that we must also devise a test that is more sensitive
to a systematic offset.  For this test, we compute the eight values of
$\delta f = f^{(r<21)}-f^{(r>21)}$ and their errors $\sigma(\delta
f)$, then find $\langle \delta f\rangle$ via a weighted (by
$1/\sigma^2(\delta f)$) average over $\delta f$ values, and the error
on this quantity, $\sigma(\langle \delta f\rangle)$.  The result is
that the weighted average $\langle\delta f\rangle=-0.13$, with an
error $\sigma(\langle \delta f\rangle)=0.10$.  Consequently, the
discrepancy is just over $1\sigma$, and therefore not statistically
significant. 

 As shown,
the constraints for the blue galaxies are weaker 
than for the red galaxies.  When we average over all source samples
for a given luminosity bin for blue galaxies, we find results that are
negative for L3--L5 but consistent with zero at the $1-1.5\sigma$
level, and for L6, that are positive but again consistent with zero at
the $1\sigma$ level.  For
red galaxies, the results are somewhat more complicated.  The measured
value of $f-f_{45}$ appears to roughly increase with luminosity,
which can be explained in several ways, as we will show.  We do not show results averaged over lens
luminosities, because the results for $f-f_{45}$ must first be
related to predictions for $f-f_{45}$ for a given profile to extract
the measured $f_h$ for that model, and
{\it then} averaged over luminosity.  In other words, for a given
profile, the resulting $f-f_{45}$ for these radial ranges are not
the same for different lens luminosities even for the same
$f_h$ (due to variation of $r_s$ with luminosity), so an average of
$f-f_{45}$  over luminosity is meaningless.

We now relate the non-parametric
determination of $f-f_{45}$ in Table~\ref{T:fnonpar} to $f_h$
for various halo profiles.  
However, because our results for $f-f_{45}$ have such large errors, we
will not use an extremely detailed or precise model.  
The simplest model, a SIS, predicts $f/f_h=0.25$ and $f_{45}=0$ on all
scales, so that for all the results above, we can write $f_h =
(f-f_{45})/0.25$.  
We also consider the NFW and TIS models, but do not attempt to do
detailed profile fitting, which is complicated by
possible calibration errors due to use of photometric redshifts for
foregrounds.  What
matters is only an approximate value of $r_s$ for the NFW and TIS
profiles.  To get an approximate value of $r_s$ using the NFW profile, we use
best-fit masses for these lens samples from the halo model fits as
described in \cite{2005MNRAS.362.1451M} and
\cite{2005PhRvD..71d3511S}. Since the calibration for the photometric
foreground samples is uncertain, we use the same luminosity and colour 
divisions with the spectroscopic sample lenses, and neglect evolution in 
the mass-luminosity relationship with redshift (though we do include
passive luminosity evolution as noted in \S\ref{SS:lens}).  
For the given best-fit mass, we can compute the concentration
parameter at the effective redshift of these lens samples using the
relationship used in that work and in \cite{2005MNRAS.362.1451M},  
\be
c(M,z) = 10\left(\frac{M}{M_{nl}(z)}\right)^{-0.13}.
\ee
Furthermore, since for
an NFW profile the mass and virial radius (where virial mass here is the
mass within the radius at which the density is 180 times the mean
density, or 54 times the critical density with $\Omega_m=0.3$)
are related, the halo mass and redshift alone determine the scale radii for these luminosity bins.  
We then used the
profile-dependent forms for $f-f_{45}$ in Figs.~\ref{F:fradTIS} and \ref{F:fradNFW} 
to redo the nonparametric determination of $f-f_{45}$ described above
dividing by $(f-f_{45})/f_h$ for our models in each bin, yielding a determination
of $f_h$ for these two profiles that includes non-Gaussianity when
averaging over these radial bins.
For the TIS profile, we fit \ds{} for the profiles to obtain scale
radii only (since the amplitude is a nuisance parameter).
Table~\ref{T:fmodel} shows relevant model parameters for the SIS
profiles, for the NFW model ($M_{180}$ and $r_s$),
and for the TIS profile ($r_s$), in addition to the resulting $f_h$
from the non-Gaussian averaging method.  We note that the scale radii
show the expected trend of increasing with luminosity, and that the
best-fit masses are consistent with those from
\cite{2005PhRvD..71d3511S} for each bin when we average over the
results for each colour.

First, we consider the results for blue lenses in Table~\ref{T:fmodel}.  With the SIS profile,
L3--L5 each have $f_h \sim 1\sigma$ negative, and L6 has $f_h \sim 1\sigma$
positive, yielding a net result of $f_h=-1.1\pm 0.6$, so almost
$2\sigma$ negative.  We note that the results for each luminosity bin
are statistically consistent at the $1\sigma$ level, and that the
result averaged over luminosity is not equal to that with $1/\sigma^2$
weighting because it includes the non-Gaussianity of the error
distributions.  With the NFW profile, the measured $f_h$ values are
larger in magnitude than with the SIS, as expected due to the lower
predicted $(f-f_{45})/f_h$ for this model, and the averaged result is
negative but consistent with zero; the same is true for the TIS.

For red lenses, the assumption of an SIS profile gives a net value of
$f_h=-0.06\pm 0.19$, so consistent with no halo ellipticity.  However,
the NFW and TIS profiles, with lower $(f-f_{45})/f_h$
(even going slightly negative for the TIS), yield positive
luminosity-averaged results that are quite similar to each other, $0.60\pm 0.38$
(NFW) and $0.57\pm 0.41$ (TIS), but again consistent with zero.  We note
that the errors for the NFW and TIS profiles are larger than for the
SIS profile; this is likely due to the fact that the predicted
$(f-f_{45})/f_h$ values are so much lower and vary with radius, so
since we divide by this small number on large scales, these scales do not
dominate as much as they do for the SIS model, where the low measured
values of $f_h$ tend to dominate the averaging process and lead to a
small result with small errors.
For red galaxies, the results in different
luminosity bins appear to be statistically inconsistent; use of the $\chi^2$
test for sample consistency in Eq.~\ref{E:chi2test} yields $\chi^2=8$
for 3 degrees of freedom, or $p(>\Delta\chi^2)=0.05$.  Hence, while they are not
definitively discrepant, there is still a suggestion
of either increasing halo ellipticity, or increasing alignment of halo
light and mass with luminosity.  

While one might be concerned about the model-dependence of these NFW
and TIS
predictions, we have verified that for red galaxies, changing the scale radius $r_s$ by
20 per cent in either direction changes the central value of $f_h$ by
roughly $0.5\sigma$, so the conclusions we derive from these results
are not too strongly sensitive to errors in our derivations of $r_s$.
(For NFW profiles, that change in $r_s$ corresponds to either a 20 per
cent change in $c(M,z)$ for a given halo mass, or a 25 per cent
change in the halo mass using our assumed $c(M,z)$ relation.  For red
galaxies, the central halo mass is generally known to this precision
or better, as in Table~\ref{T:fmodel}.)  For blue galaxies, the halo mass is
not as well-determined, but neither is $f-f_{45}$, so once again the
possible systematic error in $f_h$ due to uncertainty in $r_s$ is
within $1\sigma$.
Furthermore, since the results with NFW and TIS profiles agree at the
$1\sigma$ level, it seems that they are not too sensitive to the exact
form of the density profile assumed, as long as it is one that is
steeper at larger radii, with the scale radius associated with this
change increasing with luminosity as we have used here.

\begin{table*}
\caption{\label{T:fmodel} Expected values of $f_h$ averaged over radial
 ranges described in the text, and over source samples for each lens 
  sample, using SIS, NFW and TIS profiles with relevant parameters
 shown here.  Results are also shown averaged over luminosity and colour.}
\begin{tabular}{ccccccc}
\hline\hline
Lenses & \multicolumn{1}{c}{SIS profile} &
\multicolumn{3}{c}{NFW profile} & \multicolumn{2}{c}{TIS profile} \\
 & $f_h$ & $M_{180}$ ($10^{11} \hMsun$) & $r_s$ (\hkpc) &
$f_h$ & $r_s$ (\hkpc) & $f_h$ \\
\hline
L3, blue & $-1.1\pm 1.1$ & $6.1\pm 2.8$ & 16 & $-5.2_{-3.2}^{+2.7}$ &
180 & $-2.5_{-2.1}^{+1.9}$ \\
L4, blue & $-1.5\pm 1.0$ & $8.9\pm 3.4$ & 19 & $-1.7_{-2.7}^{+2.5}$ &
180 & $2.5_{-2.4}^{+2.8}$ \\
L5, blue & $-1.1\pm 1.1$ & $9\pm 8$ & 20 & $-9.1_{-5.6}^{+4.7}$ & 250
& $-4.6_{-3.0}^{+2.7}$ \\
L6, blue & $4.0_{-3.6}^{+5.4}$ & $165\pm 246$ & 81 &
$8.8_{-7.0}^{+15.0}$ & 400 & $18_{-13}^{+11}$ \\
All, blue & $-1.1\pm 0.6$ & - & - & $-1.4_{-2.0}^{+1.7}$ & - &
$-0.5\pm 1.3$ \\
\hline
L3, red & $-1.3\pm 0.5$ & $5.1\pm 2.5$ & 15 & $-5.3_{-2.8}^{+2.4}$ &
200 & $-1.4_{-1.3}^{+1.2}$ \\
L4, red & $0.06\pm 0.26$ & $13.0\pm 2.7$ & 23 & $0.3\pm 0.9$ & 250 &
$0.59\pm 0.62$ \\
L5, red & $0.31\pm 0.32$ & $67\pm 8$ & 52 & $0.40\pm 0.57$ & 350 &
$0.47\pm 0.75$ \\
L6, red & $1.2\pm 0.5$ & $311\pm 82$ & 108 & $1.7\pm 0.7$ & 600 &
$2.7\pm 1.0$ \\
All, red & $-0.06\pm 0.19$ & - & - & $0.60\pm 0.38$ & - & $0.57\pm 0.41$ \\
\hline\hline
\end{tabular}
\end{table*}

Next, we analyzed the results assuming a power-law model for $\ds$,
which is approximately true for the scales under consideration, doing
a joint
fit for the amplitude and slope of \ds, and for $f_h$ with
$f\ds-\fdsrot$.  
In our case, the amplitude and
$\alpha$ are just nuisance parameters, with the important quantity
being the best-fit value of $f_h$.   $f_h$ is slightly
correlated with $\alpha$ due 
to the fact that the  predicted value of $f-f_{45}$ can be related to a
parameter combination including $f_h$ and $\alpha$.  Best-fit values
of $f_h$ are shown in table~\ref{T:fhpowerlaw}; $\alpha$ is typically
$\approx 0.8$--0.9.  The
fits were done individually for each lens-source sample combination, and the results  were
averaged over lens or source samples assuming Gaussian errors. 
\begin{table*}
\caption{\label{T:fhpowerlaw} Results for $f_h$ from power-law profile
  fits described in the text for each combination of lens and source
  samples.  All errors shown are Gaussian 68 per cent confidence
  intervals.}
\begin{tabular}{cllll}
\hline\hline
Lenses & $r<21$ sources & $r>21$ sources & LRGs & all \\
 & \multicolumn{4}{c}{$f_h$} \\
\hline
L3, blue & $-6.1\pm 2.7$ & $-1.7\pm 2.0$ & $-3\pm 92$ & $-3.3\pm 1.6$ \\
L4, blue & $-1.5\pm 1.8$ & $-0.33\pm 0.99$ & $0.0\pm 4.2$ & $-0.6\pm 0.8$ \\
L5, blue & $-0.29\pm 0.66$ & $-3.5\pm 2.1$ & $-1.4\pm 2.0$ & $-0.7\pm 0.6$ \\
L6, blue & $13\pm 9$ & $10\pm 11$ & $-0.4 \pm 2.0$ & $0.5\pm 1.9$ \\
All blue & $-0.7\pm 0.6$ & $-1.0\pm 0.8$ & $-0.8\pm 1.3$ & $-0.8\pm 0.4$ \\
\hline
L3, red & $-0.26\pm 0.37$ & $-0.8\pm 0.3$ & $-1.0\pm 1.0$ & $-0.61\pm 0.23$ \\
L4, red & $-0.16\pm 0.22$ & $0.11\pm 0.18$ & $0.37\pm 0.36$ & $0.05\pm 0.13$ \\
L5, red & $-0.23\pm 0.22$ & $0.07\pm 0.17$ & $1.7\pm 0.7$ & $0.03\pm 0.13$ \\
L6, red & $0.01\pm 0.36$ & $0.21\pm 0.08$ & $0.88\pm 0.42$ & $0.21\pm 0.08$ \\
All red & $-0.17\pm 0.13$ & $0.13\pm 0.07$ & $0.63\pm 0.25$ & $0.10\pm 0.06$\\
\hline\hline
\end{tabular}
\end{table*}
As shown in this table, the results are similar to before in that (1)
the results for the $r<21$ sample are lower than for the other
samples, (2) for blue lenses, the results are negative for L3--L5 but
positive (with large error) for L6, and (3) for red lenses, the
results seem to imply some increase of ellipticity or of alignment of
light and mass ellipticities with luminosity.
However, as shown here, the values of $f_h$
implied for red lenses are lower than from the NFW and TIS analysis
from the non-Gaussian $f-f_{45}$ determination.  This is to be
expected from our analysis in \S\ref{SS:elliptheory}, and since the
non-power law profiles are more realistic, the results in
Table~\ref{T:fmodel} are probably more trustworthy.  We note that in
comparison with the SIS model results in Table~\ref{T:fmodel}, the
results with this power-law fit method averaged over all sources in
the last
column of Table~\ref{T:fhpowerlaw} (1) are different, and
(2) have significantly smaller errors.  Considering that the power-law
fit model has an additional free parameter (the power-law slope
$\alpha$) which should ostensibly lead to {\it larger} errors rather
than smaller ones, and that the change from $\alpha=1$ (SIS) to the best-fit
values $\alpha=0.8$--0.9 cannot account for the difference in the
results, these discrepancies require reconciliation.  However, it is 
important to keep in mind that the method used to determine $f_h$ for
Table~\ref{T:fmodel} correctly incorporates the non-Gaussianity of
the measurement, whereas the power-law fits used for
Table~\ref{T:fhpowerlaw} have quoted errors from Gaussian propagation of errors
on $\ds$ and $f\ds$; consequently, the larger errors in
Table~\ref{T:fmodel} are more likely to be correct.  Furthermore,
since the use of the non-Gaussian error distributions weights the
$f_h$ values in different radial ranges differently than the optimal
$1/\sigma^2$ weighting inherent in the $chi^2$-minimization power-law fits, it is not
surprising that the results from the two methods differ.

\subsection{Alignment with LSS}

Here we show results for the same tests using 800 \hkpc{} to 2 \hmpc\ 
in order to determine if there is any net alignment between the
ellipticities of the light and of local LSS (e.g., along filamentary
structures).  For these tests, we assume a power-law profile as is
approximately observed on those scales, and use both the non-Gaussian
determination and the power-law fit for $f_h$ that assumes Gaussianity
of the errors.

The non-Gaussian analysis, averaged over source and lens samples,
yields a value of $f_h = -0.46 \pm 0.34$ for spirals, and $-0.07\pm
0.09$ for ellipticals.  As mentioned previously, we expect that for
spirals we should see some tendency towards negative $f_h$ on large
scales, but unfortunately the errors are too large to be able to
confirm this result at this time.  For ellipticals,
\cite{2002ASPC..268..395K} showed that at least those that are the
BCGs of clusters  should have their ellipticities aligned with the cluster
ellipticity, and therefore we may expect positive $f_h$ for these
galaxies; this trend is not observed in this dataset.

The Gaussian analysis (a power-law fit to the signal for amplitude and
slope as nuisance parameters in addition to $f_h$) on these scales
yielded $f_h = -0.42 \pm 0.28$ for spirals, and $-0.01 \pm 0.05$ for
ellipticals.   Hence, both analysis methods yield results that are
similar, with a negative result that is not statistically significant
for spirals, and a null result that places a constraint on the
alignment of light with LSS on 0.8--2 \hmpc{} scales for ellipticals.

\section{Discussion}\label{S:discussion}

We can draw a number of interesting conclusions from the results in
\S\ref{S:results}.  First, several of our systematics tests
have interesting implications.  It is clear that in SDSS data, there
are several problems that lead to difficulties in the measurement.
The significant systematic shear, which seems to be related to the
modeling of the PSF or the PSF correction itself, requires us to undertake additional complications
in our analysis (the subtraction of $f-f_{45}$ instead of comparing
$f$ directly with the models).  Second, small-scale photometric problems  lead
to loss of sources preferentially along the lens major axis.

Some of the systematics themselves have interesting scientific
implications.  For example, we have a clear detection of anisotropy of
the distributions of satellites around lens galaxies, at least for the
brightest red galaxies ($-21 > M_r > -23$).  Since many of these
galaxies reside in 
clusters, this may reflect that the BCGs (i.e., our lenses) are
aligned with the major 
axis of clusters, as found by \cite{1982A&A...107..338B} with 44 Abell
clusters out to $z=0.1$, \cite{1999ApJ...519...22F} with poor, low
redshift clusters, \cite{2000ApJ...543L..27W} in 3 dimensions for the
Virgo cluster, and \cite{2002ASPC..268..395K} using $\sim 300$ clusters
in SDSS data over a large range in redshift, $0.04<z<0.5$.   This
effect may be explained in terms of anisotropic infall into the
cluster potential well along filaments. An upcoming work
will explore this finding in more detail.

Another interesting
systematic is the anisotropy of sources due to magnification bias,
of which hints were seen in the results, but which was not the focus of this work;
future works with higher $S/N$ on the anisotropy of lensing signal and
magnification bias, and lower contamination by physically associated
sources, should explore the relationship between the two in 
more detail, since they must be related for consistency to the average dark
matter halo profiles and alignment of light and mass ellipticities,
and may provide another measure of $f_h$ that will allow for a more
precise measurement of halo ellipticity.

Our results for halo ellipticity are not conclusive. For spirals,
we appear to have signs of anti-alignment of the ellipticities of
light and halo on a 1--$2\sigma$ level (when averaged over luminosity)
depending on the model used; for
ellipticals, we see signs of a progression, with $e_h/e_g$ increasing
with luminosity.  If we trust the many theoretical predictions of dark
matter halo ellipticity (which have been verified by many different
types of simulations, both N-body and hydrodynamic), then we must
conclude that the trend we see reflects a change in the average
alignment of the halo and light ellipticities with luminosity. 

 We
emphasize results with the non-Gaussian averaging over radial ranges
rather than the power-law fits that ignore non-Gaussianity in the
error distributions, and we emphasize those results using the NFW and
TIS models rather than the SIS or other power-law models for
$(f-f_{45})/f_h$. Our 
preference stems from the fact that, while our results in \cite{2005MNRAS.362.1451M} and
\cite{2005PhRvD..71d3511S} show that \ds{} may appear to be a power-law on
all scales shown there, 20 \hkpc{} to 2 \hmpc{}, according to the halo
model that is used for the fits to the signal, the signal due to the
central halo itself is fit more accurately with an NFW profile (or
other truncated profile that falls below the power-law on several
hundred \hkpc{} and larger scales), with the balance of the signal on
those larger scales coming from satellites within the same halo for
those galaxies that are in a group or cluster.  If one assumes that
the central halo itself has a power-law density profile, then tries to
add in the signal from other galaxies in the group/cluster, the
predicted signal becomes too large to fit the data on large
scales.  Because of the change of the density profile with radius in
these more realistic NFW/TIS models, the predicted signal anisotropy
on the largest scales probed in this paper ($\sim 300$ \hkpc) is
significantly smaller than the predicted anisotropy for a power-law
profile (with larger error), leading to a significantly larger measured value of
$f_h$ using these profiles.  We have shown that our results are not
very sensitive to even a 20 per cent change in the measured $r_s$, and are
not too different for the NFW or TIS, so they
are not too dependent on the parameters of the density
profile as long as that density profile roughly matches that of the
expected one according to the halo model fits, with a scale-dependent
power-law exponent.

There is only one previous measurement of halo ellipticity using weak lensing,
\cite{2004ApJ...606...67H}.  In that paper, the result was that
$f_h = 0.77_{-0.21}^{+0.18}$.  Our results, while more complicated
due to the apparent luminosity dependence, do not seem at first glance to agree
with this result. 
However, we must first ensure that we are comparing
the appropriate things, since these two papers use different data and
methodology.   

There are several differences in the data used for these papers.  The
SDSS data has redshifts, or at least photometric redshifts with
$\sigma(z)\sim 0.04$, for all the lenses used;
\cite{2004ApJ...606...67H} uses data from RCS, with redshift
distributions only for both lenses and sources.  Furthermore, since
their lenses have 
$19<R<21$, and mean redshift of 0.35, they correspond mainly to
our L4--L6.  
In this work, we use colour information to approximately separately 
ellipticals
and spirals;  \cite{2004ApJ...606...67H} lack that capability,
so their results are for a mixture of the two.  One might expect that
if ellipticals have light and mass ellipticities aligned, and spirals
have them anti-aligned, then \cite{2004ApJ...606...67H} should have
seen a null result for the mixture, but this is not the case.  Indeed,
if we average our results (including non-Gaussianity) from
Table~\ref{T:fmodel} over luminosities L4--L6 and over colour, we find
$f_h=0.02\pm 0.21$ for the SIS because it does not take into account
predicted radial-dependence of the results, or $f_h=0.73\pm 0.39$ for
the NFW profile and $0.86\pm 0.43$ for the TIS (note that these values
are larger than the results averaged over L3--L6 because L3 gives a
significantly negative value, and they are positive because the signal
for red galaxies entirely dominates over that of the blue galaxies).  The latter two
results do not actually contradict the result from
\cite{2004ApJ...606...67H}, so even though we cannot say they constitute a
definitive measurement of halo ellipticities, we do not have to worry
about a significant discrepancy either.  However, we note that this
procedure of averaging the $f_h$ values over colour and luminosity,
taking into account non-Gaussianity of error distributions, may not
give the same answer as the method used in \cite{2004ApJ...606...67H},
for which the signals themselves were all averaged and the maximum
likelihood method was applied to the average signal.  Furthermore, the
maximum likelihood method itself is susceptible to bias due to
systematic effects
such as the intrinsic ellipticity-density alignment that was observed
to be present at some level (but taken into account by a correction
factor) in this work.
\cite{2004ApJ...606...67H} has a mean lens redshift of 0.35 and mean
source redshift of 0.53, whereas this work has a lower mean lens
redshift for all luminosities (though the brightest bin is close to
that of \citealt{2004ApJ...606...67H}) and lower mean source redshift for
all but the LRG sample.  Therefore, if there is some evolution in
galaxy ellipticity with redshift, which has been proposed for cluster
ellipticities based on simulations
in \cite{2005ApJ...618....1H}, with higher redshift clusters having higher
ellipticity, then one might have expected a larger detection in
\cite{2004ApJ...606...67H} than here.  

The two papers also feature different analysis methods.  As argued in an appendix of \cite{2004ApJ...606...67H},
they do not expect large systematic shear, though no measurement of
$f_{45}$ was presented to show that this is the case.  However, systematic shear can only cause an underestimate
of $f$, not an overestimate.  \cite{2004ApJ...606...67H}  computed
the theoretical shear for a TIS model, then introduced anisotropy into the
  shear (i.e., not into $\kappa$ and then into the shear via
derivation of the full 2-dimensional potential as done here) assuming
that the anisotropy is equal to that in a SIS.\footnote{H. Hoekstra,
  e-mail communication.}  Since we have seen that the anisotropy for
TIS approaches that of a SIS on small scales, but decreases with
scale, the derivation of $f_h=0.77_{-0.21}^{+0.18}$ is actually an
under-estimate, since it assumes the SIS prediction even on larger
scales.   The fact that the SIS prediction for $f/f_h$ was
used means that rather than compare our NFW/TIS predictions against
theirs, we must compare our SIS prediction of $f_h=0.02\pm 0.21$
against the result of \cite{2004ApJ...606...67H}, giving a $2.5\sigma$
discrepancy if we take the errors at face value.  The
\cite{2004ApJ...606...67H} paper derives errors on $f_h$ from the
maximum likelihood method, using the $\Delta\chi^2$ values.
While this method typically gives reasonable confidence intervals even
for non-Gaussian error distributions, the different error derivation
between the two results nonetheless means that the apparent
$2.5\sigma$ discrepancy should not be considered a cause for alarm.

It is clear from this discussion that there are numerous differences 
between the two works, making the cause for the discrepancy
uncertain, the primary contenders being systematic contamination and 
physical differences of the samples used in the analysis. 
It is, however, also clear 
that we do not find a statistically
significant alignment between light and dark matter.
This may be expected from a theoretical perspective, since 
the presence of baryons tends to round the dark matter halos
\citep{2004ApJ...611L..73K}. 

Future work to measure halo ellipticity with galaxy-galaxy weak lensing
in the SDSS should have several
focuses.  First, with time there will be more data, allowing us to increase the signal to noise of these
measurements.  
Second, once there is more data, we will have the
ability to characterize the systematics, such as systematic shear and the contamination by
intrinsic alignments, better.  Finally, we hope in the near future to
have a catalog of clusters with ellipticities.  Since we are finding
hints of a positive detection of
halo ellipticity in L6, the most massive lens sample, we hope for more 
definitive constraints using clusters as lenses, since they are
more massive and the signal-to-noise on their lensing signal is
higher than for galaxies.

In conclusion, while we have not made a definitive detection of dark
matter halo ellipticity using SDSS data, this work  has made
several contributions to future efforts in this field.  We have 
identified
some of the contaminants to such a
measurement that may exist in all datasets, and therefore should be considered
in future analyses in this field.  We have found suggestions
of an anti-alignment of light ellipticities with halo ellipticity on the $1$--$2\sigma$
level for spiral galaxies (averaged over luminosity), and a suggestion
that the light and mass ellipticity alignment in ellipticals is a function of
galaxy luminosity, ranging from consistent with zero up to
a $2\sigma$ alignment in L6.  Finally, we have attempted to find
correlations between the ellipticities of light and LSS on 0.8-2
\hmpc{} scales, and have placed constraints on such a correlation for
ellipticals; results for spirals had fairly large errors so, while the
results were negative in accordance with predictions, there is no
clear detection.  Future efforts with more
data should allow us to make more definitive statements about these effects.

\section*{Acknowledgements}

R.M. is supported by an NSF Graduate Research Followship.  C.H. is
supported through NASA grant NGT5-50383. 
U.S. is supported by Packard Foundation, NASA NAG5-11489
and NSF CAREER-0132953.  We thank the referee for many useful
suggestions in improving the content and presentation of results.

Funding for the creation and distribution of the SDSS Archive has been
provided by the Alfred P. Sloan Foundation, the Participating
Institutions, the National Aeronautics and Space Administration, the
National Science Foundation, the U.S. Department of Energy, the
Japanese Monbukagakusho, and the Max Planck Society. The SDSS Web site
is http://www.sdss.org/. 

The SDSS is managed by the Astrophysical Research Consortium (ARC) for
the Participating Institutions. The Participating Institutions are The
University of Chicago, Fermilab, the Institute for Advanced Study, the
Japan Participation Group, The Johns Hopkins University, the Korean
Scientist Group, Los Alamos National Laboratory, the
Max-Planck-Institute for Astronomy (MPIA), the Max-Planck-Institute
for Astrophysics (MPA), New Mexico State University, University of
Pittsburgh, University of Portsmouth, Princeton University, the United
States Naval Observatory, and the University of Washington.


\begin{thebibliography}{}

\bibitem[\protect\citeauthoryear{{Abazajian} et~al.}{{Abazajian} et~al.}{2003}]{2003AJ....126.2081A}
{Abazajian} K. et~al., 2003, \aj, 126, 2081

\bibitem[\protect\citeauthoryear{{Abazajian} et~al.}{{Abazajian} et~al.}{2004}]{2004AJ....128..502A}
{Abazajian} K. et~al., 2004, \aj, 128, 502

\bibitem[\protect\citeauthoryear{{Abazajian} et~al.}{{Abazajian} et~al.}{2005}]{2005AJ....129.1755A}
{Abazajian} K.  et~al.,  2005, \aj, 129, 1755

\bibitem[\protect\citeauthoryear{{Bernstein} \& {Jarvis}}{{Bernstein} \&
  {Jarvis}}{2002}]{2002AJ....123..583B}
{Bernstein} G.~M.,  {Jarvis} M.,  2002, \aj, 123, 583

\bibitem[\protect\citeauthoryear{{Binggeli}}{{Binggeli}}{1982}]{1982A&A...107..338B}
{Binggeli} B.,  1982, \aap, 107, 338

\bibitem[\protect\citeauthoryear{{Blanton} et.~al.}{{Blanton}
  et~al.}{2003a}]{2003AJ....125.2348B}
{Blanton} M.~R.,  et~al.,  2003a, \aj,
  125, 2348

\bibitem[\protect\citeauthoryear{{Blanton} et.~al.}{{Blanton}
    et~al.}{2003b}]{2003ApJ...592..819B}
{Blanton} M.~R., et~al., 2003b, \apj, 592, 819

\bibitem[\protect\citeauthoryear{{Bliss}}{{Bliss}}{1935a}]{bliss1}
{Bliss} C.~I.,  1935a, Ann. Appl. Biol., 22, 134

\bibitem[\protect\citeauthoryear{{Bliss}}{{Bliss}}{1935b}]{bliss2}
{Bliss} C.~I.,  1935b, Ann. Appl. Biol., 22, 307

\bibitem[\protect\citeauthoryear{{Brainerd}}{{Brainerd}}{2005}]{2005ApJ...628L.101B}  
{Brainerd} T.~G.,  2004, \apjl, 628L, 101

\bibitem[\protect\citeauthoryear{{Brainerd} et~al.}{{Brainerd}
  et~al.}{1996}]{1996ApJ...466..623B}
{Brainerd} T.~G.,  {Blandford} R.~D.,    {Smail} I.,  1996, \apj,
466, 623

\bibitem[\protect\citeauthoryear{{Carroll} et~al.}{{Carroll}
  et~al.}{1992}]{1992ARA&A..30..499C}
{Carroll} S.~M.,  {Press} W.~H.,    {Turner} E.~L.,  1992, \araa, 30, 499

\bibitem[\protect\citeauthoryear{{Coil} et.~al.}{{Coil}
  et~al.}{2004}]{2004ApJ...609..525C}
{Coil} A.~L. et~al.,  2004, \apj, 609,
  525

\bibitem[\protect\citeauthoryear{{Cooray} \& {Sheth}}{{Cooray} \&
  {Sheth}}{2002}]{2002PhR...372....1C}
{Cooray} A.,  {Sheth} R.,  2002, \physrep, 372, 1

\bibitem[\protect\citeauthoryear{{Davis} et.~al.}{{Davis} et.~al.}{2003}]{2003SPIE.4834..161D}
{Davis} M. et~al.,  2003, in
  Proceedings of the SPIE, Volume 4834, pp. 161-172 

\bibitem[\protect\citeauthoryear{{Davis} et~al.}{{Davis}
  et~al.}{2004}]{2004astro.ph..8344D}
{Davis} M.,  {Gerke} B.~F.,    {Newman} J.~A.,  2004, 
preprint (astro-ph/0408344)

\bibitem[\protect\citeauthoryear{{Dubinski} \& {Carlberg}}{{Dubinski} \&
  {Carlberg}}{1991}]{1991ApJ...378..496D}
{Dubinski} J.,  {Carlberg} R.~G.,  1991, \apj, 378, 496

\bibitem[\protect\citeauthoryear{{Eisenstein} \& {Hu}}{{Eisenstein} \&
  {Hu}}{1998}]{1998ApJ...496..605E}
{Eisenstein} D.~J.,  {Hu} W.,  1998, \apj, 496, 605

\bibitem[\protect\citeauthoryear{{Eisenstein} et~al.}{{Eisenstein} et~al.}{2001}]{2001AJ....122.2267E}
{Eisenstein} D.~J. et~al.,  2001, \aj, 122, 2267

\bibitem[\protect\citeauthoryear{{Fieller}}{{Fieller}}{1954}]{fieller}
{Fieller} E.~C.,  1954, J.~R.~Stat.~Soc.~B, 16, 175

\bibitem[\protect\citeauthoryear{{Fischer} et~al.}{{Fischer} et~al.}{2000}]{2000AJ....120.1198F}
{Fischer} P. et~al., 2000, \aj, 120,  1198

\bibitem[\protect\citeauthoryear{{Fuller} et~al.}{{Fuller}
  et~al.}{1999}]{1999ApJ...519...22F}
{Fuller} T.~M.,  {West} M.~J.,    {Bridges} T.~J.,  1999, \apj, 519, 22

\bibitem[\protect\citeauthoryear{{Fukugita} et~al.}{{Fukugita} et~al.}{1996}]{1996AJ....111.1748F}
{Fukugita} M.,  {Ichikawa} T.,  {Gunn} J.~E.,  {Doi} M.,  {Shimasaku} K.,
  {Schneider} D.~P.,  1996, \aj, 111, 1748

\bibitem[\protect\citeauthoryear{{Gunn} et~al.}{{Gunn} et~al.}{1998}]{1998AJ....116.3040G}
{Gunn} J.~E. et~al., 1998, \aj,
  116, 3040

\bibitem[\protect\citeauthoryear{{Guzik} \& {Seljak}}{{Guzik} \&
  {Seljak}}{2002}]{2002MNRAS.335..311G}
{Guzik} J.,  {Seljak} U.~.,  2002, \mnras, 335, 311

\bibitem[\protect\citeauthoryear{{Hawley} \& {Peebles}}{{Hawley} \&
  {Peebles}}{1975}]{1975AJ.....80..477H}
{Hawley} D.~L.,  {Peebles} P.~J.~E.,  1975, \aj, 80, 477

\bibitem[\protect\citeauthoryear{{Hirata} \& {Seljak}}{{Hirata} \&
  {Seljak}}{2003}]{2003MNRAS.343..459H}
{Hirata} C.,  {Seljak} U.,  2003, \mnras, 343, 459

\bibitem[\protect\citeauthoryear{{Hoekstra} et.~al.}{{Hoekstra} et~al.}{2003}]{2003MNRAS.340..609H}
{Hoekstra} H.,  {Franx} M.,  {Kuijken} K.,  {Carlberg} R.~G.,    {Yee}
  H.~K.~C.,  2003, \mnras, 340, 609

\bibitem[\protect\citeauthoryear{{Hoekstra}, {Yee} \& {Gladders}}{{Hoekstra}
  et~al.}{2004}]{2004ApJ...606...67H}
{Hoekstra} H.,  {Yee} H.~K.~C.,    {Gladders} M.~D.,  2004, \apj, 606, 67

\bibitem[\protect\citeauthoryear{{Hogg} et~al.}{{Hogg} et~al.}{2001}]{2001AJ....122.2129H}
{Hogg} D.~W.,  {Finkbeiner} D.~P.,  {Schlegel} D.~J.,    {Gunn} J.~E.,  2001,
  \aj, 122, 2129

\bibitem[\protect\citeauthoryear{{Holmberg}}{{Holmberg}}{1969}]{1969ArAst...5..305H}
{Holmberg} E.,  1969, Arxiv Astron., 5, 305

\bibitem[\protect\citeauthoryear{{Hopkins} et~al.}{{Hopkins}
  et~al.}{2005}]{2005ApJ...618....1H}
{Hopkins} P.~F.,  {Bahcall} N.~A.,    {Bode} P.,  2005, \apj, 618, 1

\bibitem[\protect\citeauthoryear{{Hudson} et~al.}{{Hudson} et~al.}{1998}]{1998ApJ...503..531H}
{Hudson} M.~J.,  {Gwyn} S. D.~J.,  {Dahle} H.,    {Kaiser} N.,  1998, \apj,
  503, 531

\bibitem[\protect\citeauthoryear{{Ivezi{\' c}} et~al.}{{Ivezi{\' c}} et~al.}{2004}]{2004AN....325..583I}
{Ivezi{\' c}} {\v Z}. et~al., 2004, Astronomische Nachrichten, 325, 583

\bibitem[\protect\citeauthoryear{{Kazantzidis} et~al.}{{Kazantzidis}
  et~al.}{2004}]{2004ApJ...611L..73K}
{Kazantzidis} S.,  {Kravtsov} A.~V.,  {Zentner} A.~R.,  {Allgood} B.,  {Nagai}
  D.,    {Moore} B.,  2004, \apjl, 611, L73


\bibitem[\protect\citeauthoryear{{Keeton}}{{Keeton}}{2001}]{2001astro.ph..2340K}
  {Keeton} C.~R.,  2001, preprint (astro-ph/0102340)

\bibitem[\protect\citeauthoryear{{Kim} et~al.}{{Kim}
  et~al.}{2002}]{2002ASPC..268..395K}
{Kim} R.~S.~J.,  {Annis} J.,  {Strauss} M.~A.,    {Lupton} R.~H.,  2002, in ASP
  Conf. Ser. 268, p. 395

\bibitem[\protect\citeauthoryear{{Kleinheinrich}
    et~al.}{{Kleinheinrich} et~al.}{2005}]{2005A&A...439..513K} 
{Kleinheinrich} M. et~al.,  2005, \aap, 439, 513

\bibitem[\protect\citeauthoryear{{Libeskind} et~al.}{{Libeskind}
  et~al.}{2005}]{2005MNRAS.363..146L}
{Libeskind} N.~I.,  {Frenk} C.~S.,  {Cole} S.,  {Helly} J.~C.,  {Jenkins} A.,
  {Navarro} J.~F.,    {Power} C.,  2005, \mnras, 363, 146

\bibitem[\protect\citeauthoryear{{Lupton}, et~al.}{{Lupton} et~al.}{2001}]{2001ASPC..238..269L}
{Lupton} R.~H.,  {Gunn} J.~E.,  {Ivezi{\' c}} Z.,  {Knapp} G.~R.,  {Kent} S.,
   {Yasuda} N.,  2001, in ASP Conf. Ser. 238, p. 269

\bibitem[\protect\citeauthoryear{{MacGillivray} et~al.}{{MacGillivray} et~al.}{1982}]{1982MNRAS.198..605M}
{MacGillivray} H.~T.,  {Dodd} R.~J.,  {McNally} B.~V.,    {Corwin} H.~G.,
  1982, \mnras, 198, 605

\bibitem[\protect\citeauthoryear{{Madgwick} et.~al.}{{Madgwick} et~al.}{2003}]{2003ApJ...599..997M}
{Madgwick} D.~S. et~al.,  2003, \apj, 599, 997

\bibitem[\protect\citeauthoryear{{Mandelbaum}, et.~al.}{{Mandelbaum}
  et~al.}{2005a}]{2005MNRAS.361.1287M}
{Mandelbaum} R.,  et~al.,
  2005a, \mnras, 361, 1287

\bibitem[\protect\citeauthoryear{{Mandelbaum} et.~al.}{{Mandelbaum} 
  et~al.}{2005b}]{2005MNRAS.362.1451M}
{Mandelbaum} R.,  {Tasitsiomi} A.,  {Seljak} U.,  {Kravtsov} A.~V.,
  {Wechsler} R.~H.,  2005b, \mnras, 362, 1451

\bibitem[\protect\citeauthoryear{{Mandelbaum} et~al.}{{Mandelbaum}
    et~al.}{2006}]{2006MNRAS.367..611M} {Mandelbaum} R., {Hirata} C.,
    {Ishak} M., {Seljak} U., {Brinkmann} J., 2006, \mnras, 367, 611

\bibitem[\protect\citeauthoryear{{McKay} et~al.}{{McKay} et~al.}{2001}]{2001astro.ph..8013M}
{McKay} T.~A. et~al.,  2001, preprint (astro-ph/0108013)

\bibitem[\protect\citeauthoryear{{Milgrom}}{{Milgrom}}{1983}]{1983ApJ...270..3%
65M}
{Milgrom} M.,  1983, \apj, 270, 365

\bibitem[\protect\citeauthoryear{{Natarajan} \&
    {Refregier}}{{Natarajan} \&
    {Refregier}}{2000}]{2000ApJ...538L.113N} {Natarajan} P.,
    {Refregier} A., 2000, \apjl, 538, L113

\bibitem[\protect\citeauthoryear{{Navarro} et~al.}{{Navarro}
  et~al.}{1996}]{1996ApJ...462..563N}
{Navarro} J.~F.,  {Frenk} C.~S.,    {White} S.~D.~M.,  1996, \apj, 462, 563+

\bibitem[\protect\citeauthoryear{{Navarro} et~al.}{{Navarro}
  et~al.}{2004}]{2004ApJ...613L..41N}
{Navarro} J.~F.,  {Abadi} M.~G.,    {Steinmetz} M.,  2004, \apjl, 613, L41

\bibitem[\protect\citeauthoryear{{Padmanabhan} et.~al.}{{Padmanabhan}
    et.~al.}{2005}]{2005MNRAS.359..237P}
{Padmanabhan} N.  et~al., 2005, \mnras, 359, 237

\bibitem[\protect\citeauthoryear{{Peacock} \& {Dodds}}{{Peacock} \&
  {Dodds}}{1996}]{1996MNRAS.280L..19P}
{Peacock} J.~A.,  {Dodds} S.~J.,  1996, \mnras, 280, L19

\bibitem[\protect\citeauthoryear{{Pier} et~al.}{{Pier} et~al.}{2003}]{2003AJ....125.1559P}
{Pier} J.~R.,  {Munn} J.~A.,  {Hindsley} R.~B.,  {Hennessy} G.~S.,  {Kent}
  S.~M.,  {Lupton} R.~H.,    {Ivezi{\' c}} {\v Z}.,  2003, \aj, 125, 1559

\bibitem[\protect\citeauthoryear{{Richards} et~al.}{{Richards} et~al.}{2002}]{2002AJ....123.2945R}
{Richards} G.~T. et~al., 2002, \aj, 123,
  2945

\bibitem[\protect\citeauthoryear{{Sackett}}{{Sackett}}{1999}]{1999ASPC..182..393S}
{Sackett} P.~D.,  1999, in ASP Conf. Ser. 182, p. 393.

\bibitem[\protect\citeauthoryear{{Sales} \& {Lambas}}{{Sales} \&
  {Lambas}}{2004}]{2004MNRAS.348.1236S}
{Sales} L.,  {Lambas} D.~G.,  2004, \mnras, 348, 1236

\bibitem[\protect\citeauthoryear{{Sanders}}{{Sanders}}{1986}]{1986MNRAS.223..539S}
  {Sanders} R.~H.,  1986, \mnras, 223, 539

\bibitem[\protect\citeauthoryear{{Sanders} \& {McGaugh}}{{Sanders} \&
  {McGaugh}}{2002}]{2002ARA&A..40..263S}
{Sanders} R.~H.,  {McGaugh} S.~S.,  2002, \araa, 40, 263

\bibitem[\protect\citeauthoryear{{Seljak} et~al.}{{Seljak} et~al.}{2005a}]{2005PhRvD..71d3511S}
{Seljak} U. et~al., 2005a, \prd, 71, 043511

\bibitem[\protect\citeauthoryear{{Seljak} et~al.}{{Seljak}
    et~al.}{2005b}]{2005PhRvD..71j3515S} {Seljak} U., et~al., 2005b,
  \prd, 71, 103515

\bibitem[\protect\citeauthoryear{{Sheldon} et~al.}{{Sheldon} et~al.}{2004}]{2004AJ....127.2544S}
{Sheldon} E.~S. et~al.,  2004, \aj, 127, 2544

\bibitem[\protect\citeauthoryear{{Smith} et~al.}{{Smith} et~al.}{2001}]{2001ApJ...551..643S}
{Smith} D.~R.,  {Bernstein} G.~M.,  {Fischer} P.,    {Jarvis} M.,  2001, \apj,
  551, 643

\bibitem[\protect\citeauthoryear{{Smith} et~al.}{{Smith} et~al.}{2002}]{2002AJ....123.2121S}
{Smith} J.~A.  et~al.,  2002, \aj, 123, 2121

\bibitem[\protect\citeauthoryear{{Stoughton} et~al.}{{Stoughton}, et~al.}{2002}]{2002AJ....123..485S}
{Stoughton} C. et~al., 2002, \aj, 123, 485

\bibitem[\protect\citeauthoryear{{Strateva} et~al.}{{Strateva} et~al.}{2001}]{2001AJ....122.1861S}
{Strateva} I.  et~al.,  2001, \aj, 122, 1861

\bibitem[\protect\citeauthoryear{{Strauss} et~al.}{{Strauss} et~al.}{2002}]{2002AJ....124.1810S}
{Strauss} M.~A.  et~al.,  2002, \aj, 124, 1810

\bibitem[\protect\citeauthoryear{{West} \& {Blakeslee}}{{West} \&
  {Blakeslee}}{2000}]{2000ApJ...543L..27W}
{West} M.~J.,  {Blakeslee} J.~P.,  2000, \apjl, 543, L27

\bibitem[\protect\citeauthoryear{{Wright} \& {Brainerd}}{{Wright} \&
  {Brainerd}}{2000}]{2000ApJ...534...34W}
{Wright} C.~O.,  {Brainerd} T.~G.,  2000, \apj, 534, 34

\bibitem[\protect\citeauthoryear{{York} et~al.}{{York} et~al.}{2000}]{2000AJ....120.1579Y}
{York} D.~G.  et~al.,  2000, \aj, 120, 1579

\bibitem[\protect\citeauthoryear{{Zaritsky} et~al.}{{Zaritsky} et~al.}{1997}]{1997ApJ...478L..53Z}
{Zaritsky} D.,  {Smith} R.,  {Frenk} C.~S.,    {White} S.~D.~M.,  1997, \apjl,
  478, L53

\bibitem[\protect\citeauthoryear{{Zentner} et~al.}{{Zentner}
    et~al.}{2005}]{2005ApJ...629..219Z} {Zentner} A.~R., {Kravtsov}
    A.~V., {Gnedin} O.~Y., {Klypin} A.~A., 2005, \apj, 629, 219

\end{thebibliography}
\end{document}